\documentclass[pra,twocolumn,amsmath,amsfonts,amssymb,superscriptaddress,nofootinbib,showpacs]{revtex4}

\usepackage{graphicx}
\usepackage{amsmath}
\usepackage{float}
\usepackage{longtable}
\usepackage{epsfig}
\usepackage{amssymb}
\usepackage{amsfonts}
\usepackage{latexsym}
\usepackage{theorem}
\usepackage{bbm}
\usepackage{bm}
\usepackage{psfrag}

\newtheorem{define}{Definition}
\newtheorem{theo}{Theorem}
\newtheorem{lemma}[theo]{Lemma}
\newtheorem{propo}[theo]{Proposition}
\newtheorem{coro}[theo]{Corollary}

{\hfill$\blacksquare$\par\vskip12ptplus60pt}

\def\ic[#1]{{\tt [#1]}} 

\newcommand{\ket}[1]{\ensuremath{|#1 \rangle}}
\newcommand{\bra}[1]{\ensuremath{\langle #1|}}
\newcommand{\braket}[2]{\ensuremath{\langle #1 | #2 \rangle}}
\newcommand{\brakket}[3]{\ensuremath{\langle #1 | #2| #3 \rangle}}
\newcommand{\ketbra}[2]{\ensuremath{| #1 \rangle \! \langle #2 |}}
\newcommand{\kettbra}[1]{\ketbra{#1}{#1}}

\newcommand{\idty}{\ensuremath{\mathbbm{1}}}
\newcommand{\Cx}{\ensuremath{\mathbb{C}}}

\newcommand{\abs}[1]{\ensuremath{\left| #1 \right|}}
\newcommand{\norm}[1]{\ensuremath{\left\Vert #1 \right\Vert}}
\newcommand{\cbnorm}[1]{\norm{#1}_{\mathrm{cb}}}

\newcommand{\N}{\ensuremath{\mathbb{N}}}
\newcommand{\C}{\ensuremath{\mathbb{C}}}
\newcommand{\R}{\ensuremath{\mathbb{R}}}

\newcommand{\bhh}[1]{\ensuremath{\mathcal{B}(\mathcal{H}_{#1})}}
\newcommand{\bh}{\ensuremath{\mathcal{B}(\mathcal{H})}}
\newcommand{\bk}{\ensuremath{\mathcal{B}(\mathcal{K})}}
\newcommand{\bkk}[1]{\ensuremath{\mathcal{B}(\mathcal{K}_{#1})}}
\newcommand{\bhstar}{\ensuremath{\mathcal{B}_{*}(\mathcal{H})}}

\newcommand{\bhhstar}[1]{\ensuremath{\mathcal{B}_{*}(\mathcal{H}_{#1})}}
\newcommand{\bc}[1]{\ensuremath{\mathcal{B}(\mathbb{C}^{#1})}}
\newcommand{\bcstar}[1]{\ensuremath{\mathcal{B_{*}}(\mathbb{C}^{#1})}}
\newcommand{\tracenorm}[1]{\ensuremath{\| #1 \|_{1}}}

\newcommand{\tr}{\ensuremath{{\rm tr}}}

\newcommand{\id}{\ensuremath{{\rm id} \, }}

\newcommand{\hh}{{\mathcal{H}}}
\newcommand{\kk}{{\mathcal{K}}}
\newcommand{\bop}[1]{{\mathcal{B}}(#1)}  
\newcommand{\bops}[1]{{\mathcal{B}}_*(#1)}  

\newcommand{\cA}{{\mathcal{A}}}
\newcommand{\cB}{{\mathcal{B}}}
\newcommand{\cC}{{\mathcal{C}}}

\newcommand{\cF}{{\mathcal{E}}}
\newcommand{\cM}{{\mathcal{M}}}

\newcommand{\mm}{{\mathcal{M}}}

\def\chan{\Gamma}
\def\purify#1{\check{#1}}

\begin{document}

\title{Reexamination of Quantum Bit Commitment: the Possible and the Impossible}
\author{Giacomo Mauro D'Ariano}
\email{dariano@unipv.it} \affiliation{Quantum Information Theory
Group, Dipartimento di Fisica A.~Volta, Universit\`{a} di Pavia,
via Bassi 6, 27100 Pavia, Italy}
\author{Dennis Kretschmann}
\email{d.kretschmann@tu-bs.de}
\affiliation{Centre for Quantum
Computation, DAMTP, University of Cambridge, Wilberforce Road,
Cambridge CB3 0WA, United Kingdom} \affiliation{Institut f\"ur
Mathematische Physik, Technische Universit\"at Braunschweig,
Mendelssohnstra{\ss}e~3, 38106 Braunschweig, Germany}
\author{Dirk Schlingemann}
\email{d.schlingemann@tu-bs.de}
\affiliation{Institut f\"ur
Mathematische Physik, Technische Universit\"at Braunschweig,
Mendelssohnstra{\ss}e~3, 38106 Braunschweig, Germany}
\affiliation{ISI Foundation, Quantum Information Theory Unit,\\
 Viale S. Severo 65, 10133 Torino, Italy}
\author{Reinhard F.~Werner}
\email{r.werner@tu-bs.de}
\affiliation{Institut f\"ur
Mathematische Physik, Technische Universit\"at Braunschweig,
Mendelssohnstra{\ss}e~3, 38106 Braunschweig, Germany}
\date{15~Oct.~2007}

\begin{abstract}
    Bit commitment protocols whose security is based on the laws of
    quantum mechanics alone are generally held to be impossible. In
    this paper we give a strengthened and explicit proof of this
    result. We extend its scope to a much larger variety of protocols,
    which may have an arbitrary number of rounds, in which both
    classical and quantum information is exchanged, and which may
    include aborts and resets. Moreover, we do not consider the receiver to be
    bound to a fixed ``honest'' strategy, so that ``anonymous state
    protocols'', which were recently suggested as a possible way to
    beat the known no-go results are also covered. We show that any
    concealing protocol allows the sender to find a cheating strategy, which is
    universal in the sense that it works against any strategy of the receiver.
    Moreover, if the concealing property holds only approximately, the
    cheat goes undetected with a high probability, which we explicitly
    estimate. The proof uses an explicit formalization of general two
    party protocols, which is applicable to more general situations, and a new
    estimate about the continuity of the Stinespring dilation of a
    general quantum channel. The result also provides a natural
    characterization of protocols that fall outside the
    standard setting of unlimited available technology,
    and thus may allow secure bit commitment. We
    present a new such protocol whose security, perhaps
    surprisingly, relies on decoherence in the receiver's lab.
\end{abstract}


\maketitle

\tableofcontents

\section{Introduction}
    \label{sec:intro}

Bit commitment is a cryptographic primitive involving two
mistrustful parties, conventionally called Alice and Bob. Alice is
supposed to submit an encoded bit of information to Bob in such a
way that Bob has (almost) no chance to identify the bit before
Alice later decodes it for him, whereas Alice has (almost) no way
of changing the value of the bit once she has submitted it: in
technical terms, a good bit commitment protocol should be
simultaneously {\em concealing} and {\em binding}.

Bit commitment has immediate practical applications, and is also
known to be a very powerful cryptographic primitive. It was
conceived by Blum \cite{Blu83} as a building block for secure coin
tossing. Bit commitment also allows to implement secure oblivious
transfer \cite{BBC+91,Cre94,Yao95}, which in turn is sufficient to
establish secure two-party computation \cite{Kil88,CVT95}.

A standard example to illustrate bit commitment is for Alice to
write the bit down on a piece of paper, which is then locked in a
safe and sent to Bob, whereas Alice keeps the key. At a later
time, she will unveil by handing over the key to Bob. However, Bob
has a well-equipped toolbox at home and may have been able to open
the safe in the meantime. So while this scheme may offer
reasonably good practical security, it is in principle insecure.
Yet all bit commitment schemes that have wide currency today rely
on such technological constraints: not on strongboxes and keys,
but on unproven assumptions that certain computations are hard to
perform. Several such protocols have been suggested, either
computationally binding \cite{Blu83,BCC88,Hal95,HM96} or
computationally concealing \cite{Nao91,OVY92}. Cryptographers have
long known that without such technological constraints, bit
commitment (like any other interesting two-party cryptographic
primitive) cannot be securely implemented in a classical world
\cite{Kil88}.

It has therefore been a long-time challenge for quantum
cryptographers to find {\em unconditionally secure} quantum bit
commitment protocols, in which --- very much in parallel to
quantum key distribution \cite{BB84,Eke91} --- security is
guaranteed by the laws of quantum physics alone.

\subsection{Quantum Bit Commitment and the No-Go Theorem}
    \label{sec:qbc}

The first quantum bit commitment protocol is due to Bennett and
Brassard and appears in their famous 1984 quantum cryptography
paper \cite{BB84}, in a version adapted to coin tossing. In their
scheme, Alice commits to a bit value by preparing a sequence of
photons in either of two mutually unbiased bases, in a way that
the resulting quantum states are indistinguishable to Bob. The
authors show that their protocol is secure against so-called {\em
passive cheating}, in which Alice initially commits to the bit
value $k$, and then tries to unveil $1-k$ later. However, they
also prove that Alice can cheat with a more sophisticated
strategy, in which she initially prepares pairs of maximally
entangled states instead, keeps one particle of each pair in her
lab and sends the second particle to Bob. It is a direct
consequence of the EPR effect that Alice can then unveil either
bit at the opening stage by measuring her particles in the
appropriate basis, and Bob has no way to detect the difference.

Subsequent proposals for bit commitment schemes tried to evade
this type of attack by forcing the players to carry out
measurements and communicate classically as they go through the
protocol. At a 1993 conference Brassard, Cr\'epeau, Jozsa, and
Langlois presented a bit commitment protocol \cite{BCJ+93} that
was claimed and generally accepted to be unconditionally secure.

In 1996 it was then realized by Lo and Chau \cite{LC97,LC98}, and
independently by Mayers \cite{May96,May97,BCM+97} that all
previously proposed bit commitment protocols are vulnerable to a
generalized version of the EPR attack that renders the BB84
proposal insecure, a result they slightly extended to cover
quantum bit commitment protocols in general. In essence, their
proof goes as follows: At the end of the commitment phase, Bob
will hold one out of two quantum states $\varrho_k$ as proof of
Alice's commitment to the bit value $k \in \{0,1\}$. Alice holds
its purification $\psi_k$, which she will later pass on to Bob to
unveil. For the protocol to be concealing, the two states
$\varrho_k$ should be (almost) indistinguishable, $\varrho_0
\approx \varrho_1$. But Uhlmann's theorem \cite{Uhl76,NC00} then
implies the existence of a unitary transformation $U$ that
(nearly) rotates the purification of $\varrho_0$ into the
purification of $\varrho_1$. Since $U$ is localized on the
purifying system only, which is entirely under Alice's control,
Lo-Chau-Mayers argue that Alice can switch at will back and forth
between the two states, and is not in any way bound to her
commitment. As a consequence, any concealing bit commitment
protocol is argued to be necessarily non-binding.

These results still hold true when both players are restricted by
superselection rules \cite{KMP04}. So while the proposed quantum
bit commitment protocols offer good practical security on the
grounds that Alice's EPR attack is hard to perform with current
technology, none of them is unconditionally secure. Spekkens and
Rudolph \cite{SR01} extended the no-go theorem by providing
explicit bounds on the degree of concealment and bindingness that
can be achieved simultaneously in any bit commitment protocol,
some of which they showed can be saturated.

\subsection{Two Camps}
    \label{sec:camps}

In view of these negative results, subsequent research has
primarily focused on bit commitment under plausible technological
constraints, such as a limited classical \cite{CCM98,DHR+04} or
quantum \cite{DFS+05} memory, or the difficulty of performing
collective measurements \cite{Sal98}. In an alternative approach,
researchers have slightly modified the standard setting to evade
the no-go theorem: Kent \cite{Ken99,Ken05} has shown that
relativistic signalling constraints may facilitate secure bit
commitment when Alice and Bob each run two labs a (large) distance
apart and security is maintained through a continual exchange of
messages. A different variant was introduced by Hardy and Kent
\cite{HK04}, and independently by Aharonov {\it et al.}
\cite{ATV+00}: in {\em cheat-sensitive} bit commitment protocols,
both players may have the chance to cheat, but face the risk of
their fraud being detected by the adversary. Building on Kent's
original proposal \cite{Ken03}, the tradeoff between bindingness
and concealment in quantum {\em string} commitment protocols has
recently been investigated \cite{BCH+05,BCH+06,Jai05}.

At the same time, the Lo-Chau-Mayers no-go theorem
\cite{LC97,May97} is continually being challenged. Yuen and others
have repeatedly expressed doubts in Mayer's opaque paper
\cite{May97}, arguing that the no-go proof is not general enough
to exclude all conceivable quantum bit commitment protocols.
Several protocols have been proposed and claimed to circumvent the
no-go theorem (see \cite{Yue00,Yue03,Yue05,Yue07,Che01} and references
therein, as well as this account \cite{Dar02a,Dar02b} of the
controversy). These protocols seek to strengthen Bob's position
with the help of `secret parameters' or `anonymous states', so
that Alice lacks some information to cheat successfully: while
Uhlmann's theorem would still imply the existence of a unitary
cheating transformation as described above, this transformation
might be unknown to Alice.

Two camps seem to have formed, a large one comprising most of the
community, in which the impossibility of quantum bit commitment is
accepted on the basis of the Lo-Chau-Mayers proof, and a smaller
group of sceptics, which is not convinced, even though no provably
secure protocol, and hence a counterexample to the no-go result,
has surfaced so far.

It appears that much of this controversy stems from slightly
differing approaches to the problem. A good way to pinpoint the
basic disagreement is Kerckhoffs' principle, which goes back to
the 19th century military cryptographer Auguste Kerckhoffs and is
now universally embraced by cryptographers \cite{TW06,Rud02}. The
principle states that the security of a cryptographic protocol
should not rely on keeping parts of the algorithm secret. In the
words of Bruce Schneier, ``every secret creates a potential
failure point. Secrecy, in other words, is a prime cause of
brittleness --- and therefore something likely to make a system
prone to catastrophic collapse'' \cite{Sch02}. In this respect
every secret parameter chosen by the human in a cryptographic
protocol --- e.~g. a password --- is regarded as a potential
weakness. For this reason cryptographers usually think of their
algorithms as being executed by machines, whose blueprints can be
published without jeopardizing the security of the system.

Anonymous states and other secret parameters used in Yuen's
protocols are apparently regarded as a violation of Kerckhoffs'
principle, which suggests a restriction to fixed and automatizable
strategies for both players. Deviations from these strategies are
considered an attempted fraud. The Kerckhoffian security analysis
then does not hold any provisions for the case in which {\em both}
parties deviate from their `honest' strategies. Therefore
Lo-Chau-Mayers only consider the final committed state given that
Bob sticks to his publicly known strategy, since Alice's cheat
only has to work against this strategy. So while Kerckhoffs'
principle is certainly high on the list of desiderata for
cryptographic protocols, it appears that Lo-Chau-Mayers only show
that there is no bit commitment protocol {\it satisfying
Kerckhoffs' principle}, whereas the next best thing, e.g., an
anonymous state protocol might still exist.

Another possible origin for disagreement is the style of Mayers's
paper \cite{May97}, along the lines of Mark Kac's dictum ``A
demonstration convinces a reasonable man; a proof convinces a
stubborn man''\footnote{Cited after N. D. Mermin: {\em Exact Lower
Bounds for Some Equilibrium Properties of a Classical
One-Component Plasma}, Phys. Rev. {\bf 171} (1968) 272, footnote
2.}. In this sense, i.e., according to the standards of
``stubborn'' mathematics or mathematical physics, Mayers gives
merely a demonstration. Since the argument against Kerckhoffian
protocols only involves the state directly after commitment,
Mayers declares it irrelevant to formalize the class of two-party
protocols, even though an insufficiently specified domain usually
leaves a no-go ``theorem'' rather fuzzy. Other aspects of the
problem (e.g. the use of classical and quantum information
together) get a similarly rough treatment. This may be a symptom
of the ``Four Page Pest''\footnote{I.e., the disease of cramming
an argument onto four pages in PRL format, although its shortest
intelligible presentation requires more than six.}. In any case,
it appeared to us high time to convince ourselves, and hopefully
some other stubborn men, of the exact scope and status of the No
Bit Commitment statements.

\subsection{A Stronger No-Go Theorem: Overview and Outline}
    \label{sec:outline}

In this contribution we propose to resolve the bit commitment
controversy with a strengthened no-go theorem. We will give a
precise description of general two-party protocols, which we hope
no longer shows the hard work of keeping it fully explicit but
still notationally manageable. This description should also be
helpful for analyzing protocols for other tasks, involving any
number of parties. Our description of bit commitment does not
assume Kerckhoffs' principle, so that Bob is not honor bound to a
particular course of action. Nevertheless, we show that any
concealing protocol allows Alice a universal cheating strategy,
working against all strategies of Bob simultaneously. Moreover,
our result is stable against small errors, in the sense that
nearly concealing protocols allow a nearly perfect cheat, with
explicit universal error bounds. The result is based on a
continuity theorem for Stinespring's representation \cite{KSW06},
which generalizes Uhlmann's theorem from quantum states to
channels.

Our proof includes a full treatment of classical and quantum
information flow and also covers aborts and resets. It applies to
bit commitment protocols with any (finite or infinite) number of
rounds during each the commitment, holding, and opening phase. We
only require that the expected number of rounds is finite.
Moreover, the proof is not restricted to quantum systems on
finite-dimensional Hilbert spaces. The strengthened no-go theorem
shows the insecurity of all recently proposed bit commitment
protocols \cite{Yue00,Yue03,Yue05,Yue07,Che01}. A preliminary
version of the proof, restricted to single-round commitments, has
appeared in \cite{BDR05}. Our results generalize that of Ozawa
\cite{Ozawanote} and recent work by Cheung \cite{Che05}, who
showed that Alice can still cheat in protocols with secret
parameters for the simpler case of perfect concealment, and
without a full reduction. Cheung's estimates \cite{Che06} for
approximately concealing protocols depend on the dimensions of the
underlying Hilbert space, and hence cannot rule out bit commitment
protocols with high-dimensional or infinite-dimensional systems.

We also classify those protocols that fall outside the standard
setting, and thus may allow secure bit commitment. We propose a new such bit
commitment protocol whose security --- perhaps paradoxically
--- relies on decoherence in Bob's lab. Interestingly, this
protocol explores a purely quantum-mechanical effect: the
distinction between the local erasure of information and the
destruction of quantum correlations \cite{HLS+04}. Well-known
classical bit commitment protocols whose security relies on noisy
communication channels are briefly reviewed, too.

The paper is organized as follows: In Section~\ref{sec:setup} we
give a detailed description of the setup for quantum bit
commitment protocols, and list important types of protocols that
fall within our definition. This will serve to specify the domain
for the proof of the strengthened no-go theorem, which is then
presented in Section~\ref{sec:proof}. In Section~\ref{sec:energy}
we briefly describe how to extend the no-go theorem to quantum bit
commitment protocols in infinite-dimensional Hilbert spaces or
with infinitely many rounds. Section~\ref{sec:decoherence}
investigates provably secure bit commitment protocols whose
security is built on decoherence in either Alice's or Bob's lab,
or in the transmission line. We conclude with a Summary and
Discussion in Sec.~\ref{sec:discussion}. An appendix contains the
necessary background on quantum states and channels, direct sums
and quantum-classical hybrid systems.


\section{The Setup}
    \label{sec:setup}

In this section we describe the task of quantum bit commitment,
and define what a successful bit commitment protocol would have to
achieve. We have attempted not to exclude any possibilities, and
have avoided all simplifications ``without loss of generality'' at
this stage. In this way we hope to separate, more clearly than our
predecessors, the definition of bit commitment to which the
statement ``Bit commitment is impossible'' refers  and, on the
other hand,  the simplifications which we will make in the course
of the proof of this statement.

The analysis will be based solely on the principles of quantum
mechanics, including classical physics. We do not consider
relativistic signalling constraints, which are known to facilitate
secure bit commitment \cite{Ken99,Ken05}. For ease of presentation, we initially impose as a {\it finiteness condition}, that all classical messages can only take finitely many values, that all quantum systems can be described in a finite dimensional Hilbert space, and that the total number of
messages exchanged is uniformly bounded. These constraints will then be relaxed in Sec.~\ref{sec:energy}.


\subsection{Description in Plain English}
    \label{sec:english}

{\bf The Basic Task ---} Bit Commitment is a cryptographic
primitive involving two mistrustful parties, conventionally called
Alice and Bob. Alice is supposed to submit an encoded bit of
information to Bob in such a way that Bob has (almost) no chance
to identify the bit before Alice decodes it for him, and Alice has
(almost) no way of changing the value of the bit after she has
submitted it. In other words, Bob is interested in {\em binding}
Alice to some commitment, whereas Alice would like to {\em
conceal} her commitment from Bob.

{\bf Protocols and Strategies ---} A {\it protocol} first of all
regulates the  exchange of messages between Alice and Bob, such
that at every stage it is clear what type of message is expected
from the participants, although, of course, their content is not
fixed. The expected message types can be either classical or
quantum or a combination thereof, with the number of
distinguishable classical signals and the dimension of the Hilbert
spaces fixed. The type of messages can depend on classical
information generated previously. The collection of all these
instructions will be called the {\it communication interface} of
the protocol.

A particular plan for operating a local laboratory to supply the
required messages, is called a {\it strategy}. A strategy could
determine that some message sent is obtained from a measurement on
a system available in the local lab, but it could also specify the
arbitrary invention of a classical value to be sent and the fresh
preparation of an accompanying quantum system. We typically denote
Alice's strategy by $a$ and Bob's by $b$.

The second key element of the protocol specifies definite
procedures for Alice to follow if she wants to commit the bit
values $0$ or $1$, respectively. These special {\it honest
strategies} will be denoted by $a_0$ and $a_1$.

{\bf Phases of the Protocol  ---} In any commitment scheme, we can
distinguish three phases. The first is the {\it commitment phase},
in which Alice and Bob start from some (publicly known and trusted) shared quantum or classical state and go through a possibly complicated exchange of classical and quantum messages. By definition, at the end of this phase, the bit value is considered to be committed to Bob but, supposedly, concealed
from him.

Alice and Bob then might split up for a while, without further
communication. In this {\it holding phase} typically only local
operations are possible, i.e., Bob might attempt to read the
committed bit, and Alice might attempt to prepare a cheat.

Finally they get in touch again to open the commitment. In the
{\it opening phase}, Alice sends to Bob some classical or quantum information to reveal her commitment. Taking both Alice's message and his own (classical and quantum) records, Bob will then perform a suitable {\it
verification measurement}. His measurement will result in either the bit value $k \in \{0,1\}$, indicating a successful commitment, or in a failure symbol ``not ok'', indicating an attempted cheat or abort.

A typical opening consists in Alice sending to Bob the value of the bit she claims to have committed, together will all the classical or quantum information needed for Bob to check this claim against his records. The protocol might also be ended in a {\it public opening}, which requires
Alice and Bob to meet, bringing with them all quantum and
classical systems in their possession, explaining what strategies
they were using, and allowing Bob to choose arbitrary measurements
on all these systems to verify, with Alice staying on to watch.
That is, no possibility of cheating, withholding information, or
making false claims about the outcome of verification exists in a
public opening.

{\bf Conditions on Successful Protocols ---} We assume that
Alice's strategies $a_0$ and $a_1$ can be distinguished with high
probability by Bob's verification measurement: if Alice honestly played $a_k$, then Bob's measurement will result in the bit value $k$ with probability $\geq(1-\eta)$ for some (small) $\eta \geq 0$. We call such a protocol {\it
$\eta$-verifiable}, or {\it $\eta$-sound}. Since this condition
depends only on the honest strategies, it is very easy to satisfy.

We call a protocol $\varepsilon$-{\it concealing}, if Alice's
honest strategies cannot be distinguished by Bob (up to an error
$\varepsilon$) before she opens the commitment. In general, of course, the probabilities he measures while applying his protocol $b$ depend on whether Alice chooses $a_0$ or $a_1$. Here we require that no matter what
strategy $b$ Bob uses and no matter what measurement he makes,
these probabilities never differ by more than $\varepsilon$
throughout the commitment and holding phase. Note that the
concealing condition makes no statement whatsoever about other
strategies of Alice. If Alice cheats, there is usually nothing to
be concealed anyway.

\begin{figure}[htb]
    \psfrag{o}{other}
    \psfrag{s}{strategies}
    \psfrag{c}{cheat}
    \psfrag{h}{honest}
    \psfrag{m}{commitment}
    \psfrag{f}{opening}
    \psfrag{a}{$a^{\sharp}$}
    \psfrag{p}{$a^{\sharp}_{0}$}
    \psfrag{q}{$a^{\sharp}_{1}$}
    \psfrag{n}{$a_{0}$}
    \psfrag{e}{$a_{1}$}
    \psfrag{d}{$D$}
    \psfrag{u}{$U$}
    \psfrag{0}{$0$}
    \psfrag{1}{$1$}
    \includegraphics[width=\columnwidth]{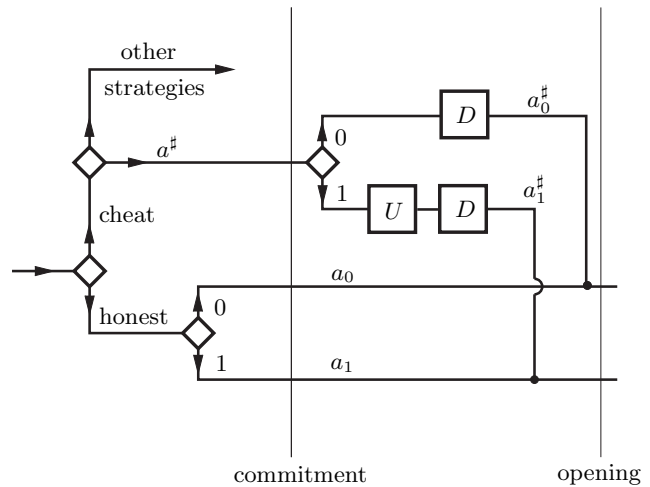}
    \caption{\label{fig:strat}{\it Alice's basic
        strategic choices. Decisions she must take are
        indicated by diamonds, some actions necessary
        for a typical cheating strategy by squares.
        The cheating strategies $a_0^\sharp$ and $a_1^\sharp$ are
        identical throughout the commitment phase, and might be equal
        to a purification of the honest strategy $a_0$. Then
        $U$ indicates a unitary cheating transformation, and
        $D$ the introduction of suitable decoherence to reverse
        purification.
        In the opening phase the cheating strategies are
        identical to their honest counterparts.}}
\end{figure}

A $\delta$-{\it cheating strategy} for Alice is a pair of
strategies $a_0^\sharp$ and $a_1^\sharp$ such that Bob cannot
distinguish $a_0$ from $a_0^\sharp$, and $a_1$ from $a_1^\sharp$
better than with a probability difference $\delta$, at any time,
including the opening phase. Of course, these conditions would be
trivially satisfied for $a_0=a_0^\sharp$ and $a_1=a_1^\sharp$.
What makes $(a_0^\sharp,a_1^\sharp)$ cheating strategies is that
Alice does not actually make the decision about the value of the
bit until after the commitment phase. That is, the strategies
$a_0^\sharp$ and $a_1^\sharp$ must be the same throughout the
commitment phase, and can only differ by local operations carried
out in the holding or opening phase. Note, however, that Alice
might have to decide from the outset that she wants to cheat,
since the strategies $a_i^\sharp$ might be quite different from
both $a_0$ and $a_1$. Fig.~\ref{fig:strat} illustrates Alice's
basic choices as she goes through the protocol. If no
$\delta$-cheating strategy exists for Alice, we call the protocol
$\delta$-binding.

The condition we impose here is much stronger than the condition
that Bob's standard verification measurements are fooled by the
cheat (perhaps with a bound on the success probability): We
require that no measurement whatsoever could detect a difference.
With a public opening rule one could even say that after the cheat
not even Alice herself could help Bob to tell the difference.
Clearly, these conditions make it very hard for Alice to cheat.
Therefore, our proof that Alice can still cheat under such
conditions automatically includes all protocols with weaker
conditions on successful cheats.

{\bf Real Time Checks for Cheating ---} It is perhaps helpful to
point out the difference between two kinds of checks on Alice's
honesty, which Bob might perform. We have granted him unlimited
technological power in the definition of $\varepsilon$-concealing.
But for running the protocol no such fantastic abilities are
required, and he will not actually do all those complicated tests.
In fact, the concealing and binding properties of the protocol
cannot be ascertained by any practical tests, but are there to be
checked theoretically by Alice and Bob on the basis of the
publicly available description of the protocol. It is on the basis
of such considerations that Alice and Bob will consent to use the
protocol in the first place.

During a single run of the protocol, Bob can employ some tests on
Alice's behavior as part of the protocol. If Bob suspects a
problem he may be entitled to calling an abort of the protocol
(clearly a classical message), and the procedure would start at
the beginning. The total number of such resets must be limited on
the grounds of bounding Alice's probability of cheating. The
possibility of such checks at run-time is the main reason why we
must consider protocols with a large number of rounds, possibly
differing from run to run.

{\bf Result ---} We will prove in Sec.~\ref{sec:nogo} that any
protocol which is $\varepsilon$-concealing allows a
$\delta$-cheating strategy for Alice, where
$\delta\leq2\sqrt\varepsilon$. These bounds coincide with those
obtained by Spekkens and Rudolph \cite{SR01} in the Kerckhoffian
setting.

As illustrated in Fig.~\ref{fig:strat}, Alice's cheating strategy
$a^\sharp$ consists in playing a purification of the honest
strategy $a_{0}$ throughout the commitment and holding phase. If
she then opts for the bit value $k=1$ instead, she will apply a
unitary operation $U$ on the purifying system, and thenceforth
follow the honest strategy $a_{1}$.


\subsection{Formal Description of Protocols}
    \label{sec:formal}

In this section we will cast the above description more explicitly
into the formalism of quantum theory. Thereby we further reduce
possible ambiguities in the statement of the problem, but also
prepare the notation for the proof.

The basic formalism of quantum theory is briefly reviewed in the
Appendix. We will generally identify systems by their observable
algebras. This has the advantage that combinations of classical
and quantum information are naturally covered: a quantum system
with Hilbert space $\hh$ is then represented by the algebra
$\bop\hh$ of operators on $\hh$, and a system characterized by a
classical parameter $x$, and has Hilbert space $\hh_x$ in that
case is described by the direct sum $\bigoplus_x\bop{\hh_x}$. A
state on such an algebra is of the form $\bigoplus_xp_x\rho_x$,
and is specified first by a probability distribution $p_x$ for the
$x$'s, and second by a collection of density operators $\rho_x$ on
$\hh_x$, which are used to compute expectations if the value of
the classical parameter is known to be $x$. Since this formalism
for handling classical information in protocols is not generally
familiar, we describe it in some more detail in Appendix A and B.

Many algebras (indexed by the nodes of the communication tree)
will appear in the description of the protocol, indicating that
with each operation the type of quantum system in the respective
lab might change completely. By choosing the lab algebras large
enough this dependence might be avoided. However, even when the
lab systems remain the same, it is helpful to keep the
distinguishing indices for keeping track of the progress of the
protocol.


\subsubsection{The communication tree}
    \label{sec:tree}

\begin{figure}[htb]
    \psfrag{A}{{\large A}}
    \psfrag{B}{{\large B}}
    \psfrag{0}{{\large 0}}
    \psfrag{x}{{\large x}}
    \psfrag{m}[cc][cc]{{\large m}}
    \psfrag{k}[cc][rr]{{\large xm}}
    \includegraphics[width=0.8\columnwidth]{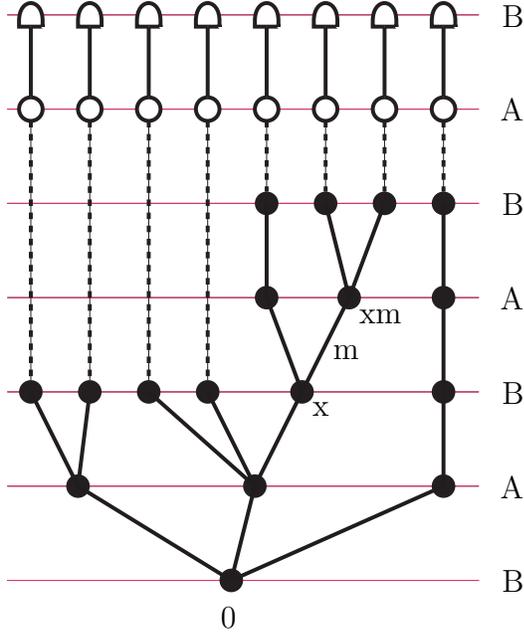}
    \caption{\label{fig:tree}{\it Example of a communication tree.
    Each node corresponds to one history of classical communications,
    with the different lines from each node representing a possible classical signal.
    The dashed lines represent the holding phase, in which no communication occurs,
    followed by the opening move (open circle) by Alice, and a measurement by Bob.}}
\end{figure}

At every stage of the protocol a certain amount of shared
classical information will have accumulated. Classical information
never gets lost, so the stages of the protocol, together with the
currently available classical information naturally form the nodes
of a tree, which we call the {\it communication tree}. An example
is depicted in Fig.~\ref{fig:tree}. Every node $x$ carries the
following information:
\begin{enumerate}
 \item Whose turn is it: Alice's or Bob's? This follows from the
 position of the node in the tree, when we assume without loss of generality,
 that Bob always starts, and from then turns alternate.
 \item What are the classical signals, which might be sent from
 this person to the other? The admissible signals form a finite
 set $M_x$ by assumption.  This set labels the branches continuing from
 this node to successor nodes which we denote by $x'=xm$,
 for $m\in M_x$.
 \item For each possible classical signal, what kind of quantum
 system is accompanying it? If the classical message is $m$, we
 take its observable algebra to be $\cM^x_m$, and assume this to be
 the full algebra of $d\times d$-matrices for some
 $d=d(x,m)<\infty$. The value $d(x,m)=1$ ($\simeq$ no accompanying quantum
 system) is a possible choice.
 \item Each node $x$ is completely characterized by the entire
 history of the classical messages exchanged between Alice and
 Bob, i.e., we can write $x=m_1m_2\cdots m_N$.
\end{enumerate}

At every node, we denote the observable algebras of Alice's and
Bob's laboratories by $\cA_x$ and $\cB_x$, respectively. These are
only partly determined by the communication interface, and depend
on the strategy, which we sometimes emphasize by writing
$\cA_x(a)$ and $\cB_x(b)$. The description of the communication
step below shows in detail how these algebras develop as one moves
along the communication tree.
Let $X_c$ denote the set of  nodes at which a commitment is
supposed to be reached. Since only local operations and the
opening phase follow, we can consider these as the leaves of the
communication tree. The joint observable algebra at that stage is
\begin{equation}\label{Obs-t}
    \bigoplus_{x\in X_c}\cA_x(a)\otimes\cB_x(b)
\end{equation}
(see the Appendix for the interpretation of direct sums). The
algebras of Alice and Bob could themselves be direct sums,
representing classical information only available to Alice and
Bob, respectively, but we do not look at this for the moment.


\subsubsection{The elementary communication step}
    \label{sec:elementary}

Now consider some node $x$, and assume that it is Alice's turn
(everything holds mutatis mutandis for Bob). We know that some
message $m\in M_x$ is expected from Alice, accompanied by a
quantum system with observable algebra $\cM_m^x$. The most general
way of doing this is a quantum operation sending states on $\cA_x$
to states on $\bigoplus_{m} \cA_{xm}\otimes\cM_m^x$. Written in
the Heisenberg picture Alice hence chooses a channel (completely
positive normalized map, cf.~Appendix)
\begin{eqnarray}
    \label{eq:elementary01a}
        T_x(a)&:&\bigoplus_{m\in M_x} \cA_{xm}(a) \otimes \cM_m^x
               \to \cA_x(a) \\
    \label{eq:elementary01b}
        \cB_{xm}(b) &=& \cM_m^x \otimes \cB_x(b).
\end{eqnarray}
Here we have added a parameter $a$ to $T_x$, to make it clear that
choosing these channels for all $x$ is precisely what defines
Alice's strategy. Note that the choice of the channel includes
that of their domain and range algebras. The channel $T_x(a)$,
together with the input state determines the probabilities for the
classical outcomes $m$. Of course, the channel could be one that
simply forces one of the results. Hence $m$ could equally well be
the result of Alice's free choice of strategy, or of a measurement
on a system she recently obtained from Bob. If $m$ is found, Alice
also splits the output system into a part $\cA_{xm}(a)$ which
Alice keeps, and the part $\cM_m^x$ she sends to Bob. This
splitting is included in the specification of $T_x(a)$. That
$\cM_m^x$ changes ownership is expressed in the above equation by
including it in Bob's algebra at the next round (i.e., $\cB_{xm}$)
as a tensor factor. At Bob's nodes everything is the same, but
since we always order tensor factors as
Alice~$\otimes$~Message~$\otimes$~Bob, the analogues of the above
equations at Bob's nodes are
\begin{eqnarray}
    \label{eq:elementary02}
    T_x(b)&:&\bigoplus_{m\in M_x} \cM_m^x\otimes\cB_{xm}(b)
               \to\cB_x(b)\\
    \cA_{xm}(a)&=&\cA_x(a)\otimes\cM_m^x.
\end{eqnarray}


\subsubsection{From commencement to commitment}
    \label{sec:commence}

We assume that Alice and Bob initially share the quantum or
classical state $\rho_{0} \mathpunct : \cA_{0} \otimes \cB_{0}
\rightarrow \Cx$. The joint state
\begin{equation}
    \label{eq:elementary03}
        \rho_c(a_{},b_{},\rho_{0}):\bigoplus_{x\in X_c}\cA_x(a)\otimes\cB_x(b) \to\Cx
\end{equation}
at commitment time is then $\rho_c(a_{},b_{},\rho_{0}) = \rho_0 \circ T(a,b)$, where
\begin{equation}
    \label{eq:elementary03b}
        T_{c}(a,b) :\bigoplus_{x\in X_c}\cA_x(a)\otimes\cB_x(b) \to\ \cA_{0} \otimes \cB_{0}
\end{equation}
is the direct sum channel that arises from the concatenation of all
the elementary step channels $T_{x}(a)$ and $T_{x}(b)$ up to
commitment time. Of course, Eq.~(\ref{eq:elementary03b}) holds
correspondingly for the shared states at all other stages of the
protocol. In particular, the final state $\rho_f(a_{},b_{},\rho_{0})$
on which Bob carries out the verification measurement arises from
the initial state $\rho_0$ by means of a quantum operation
\begin{equation}
    \label{eq:elementary03c}
        T_{f}(a,b) :\bigoplus_{x\in X_f}\cA_x(a)\otimes\cB_x(b) \to\ \cA_{0} \otimes \cB_{0} \, \,
\end{equation}
where $X_f$ is the collection of all the leaves of the
communication tree. We will assume that the initial state
$\rho_{0} \mathpunct : \cA_{0} \otimes \cB_{0} \rightarrow \Cx$ is
known to both Alice and Bob, and henceforth write
$\rho_c(a_{},b_{})$ instead of $\rho_c(a_{},b_{},\rho_{0})$, and
$\rho_f(a_{},b_{})$ instead of $\rho_f(a_{},b_{},\rho_{0})$ to
streamline the presentation.


\subsubsection{Can Bob distinguish Alice's strategies?}
    \label{sec:distinguish}

In the concealing condition, as well as in the description of
cheating strategies, it is important to decide whether Bob can
distinguish two strategies of Alice at commitment time. Clearly,
this depends only on the restriction $\rho^B_c(a_i,b)$ of the
state $\rho_c(a_i,b)$ to Bob's laboratory, which has observable
algebra $\bigoplus_x\cB_x$.

The security criterion given in Sec.~\ref{sec:english} asks for
the largest probability difference obtainable by Bob. It is
convenient to express this in a trace norm difference: the largest
difference of expectations in ``yes-no'' experiments with density
matrices $\rho_1,\rho_2$ is $\sup_F\abs{\tr(\rho_1-\rho_2)F}$,
where $F$ ranges over all so-called {\em effects} $F$ with $0\leq
F\leq\idty$. That is, the largest probability difference is
$\frac12\norm{\rho_1-\rho_2}_1$, where $\norm\cdot_1$ denotes the
trace norm. This naturally leads us to the following definition
of concealing protocols and cheating strategies:
\begin{define}
    \label{def:concealing}
        {\bf (Concealing)}\\
        We say that a protocol with a strategy pair $(a_0, a_1)$ for Alice is $\varepsilon$-concealing iff for all
        strategies $b$ of Bob
        \begin{equation}
            \label{eq:setup02}
                \norm{\rho^B_c(a_0,b)-\rho^B_c(a_1,b)}_1
                \leq 2 \, \varepsilon \, .
        \end{equation}
        When this condition holds with $\varepsilon=0$,
        we say that the protocol is {\em perfectly
        concealing}.
\end{define}
Note that one possible measuring strategy for Bob is to actually
make the measurement at an earlier time, record the result, and
send only dummy messages to Alice afterwards. So saying that two
strategies are $\varepsilon$-equivalent {\it at} some stage is the
same as saying that they are equivalent {\it up to} that stage of
the protocol. Hence the $\varepsilon$-concealing condition implies
the only apparently stronger statement that at no time during the
commitment phase Bob is able to discriminate the honest
commitments better than with probability $\varepsilon$.
\begin{define}
    \label{def:cheating}
        {\bf (Cheating)}\\
        A pair of strategies $(a_0^\sharp, a_1^\sharp)$ for Alice that coincide until after the commitment phase is called a $\delta$-cheating strategy iff
        \begin{equation}
            \label{eq:setup03}
                \norm{\rho^B_f(a_i^\sharp,b)-\rho^B_f(a_i,b)}_1\leq 2 \, \delta \, ,
        \end{equation}
        for Alice's honest strategies $(a_0, a_1)$, $i=0,1$ and all of Bob's strategies $b$.
\end{define}
Def.~\ref{def:cheating} requires a cheating strategy to work
against {\em all} of Bob's strategies --- not only against some
{\em fixed} strategy, as suggested by Kerckhoffs' Principle. We
will show in Sec.~\ref{sec:nogo} that Alice can always find such a
universally good cheating strategy. As explained in the
Introduction, this extends the no-go theorem to protocols relying
on secret parameters or ``anonymous states''. If Bob's strategy $b$ is supposed to be fixed and publicly known in Eq.~(\ref{eq:setup03}), our no-go proof will reduce in essence to the one obtained previously by Lo-Chau-Mayers \cite{LC97,LC98,May96,May97}.


\subsection{Protocols Covered by our Definition}

In this section we describe some ideas from the literature about
possible protocols, in increasing complexity. Of course, none of
them are ultimately successful. But this is in many cases not
obvious from the outset, so these ideas serve well to illustrate
the richness of two-party protocols as formalized in our scheme.


\subsubsection{The beginning}

As explained in the Introduction, the first observation concerning
quantum bit commitment was made in the classic paper of Bennett
and Brassard on quantum cryptography \cite{BB84}. In this basic
scenario the commitment phase has only one round, in which Alice
prepares one of two orthogonal Bell states
$\psi_0,\psi_1\in\hh_A\otimes\hh_B$. These have the same
restriction on Bob's system, so the protocol is perfectly
concealing. But they are also connected by a unitary on Alice's
side (as all maximally entangled states are), and this unitary
constitutes her {\em sneak flip} cheating strategy, which under these
circumstances also works perfectly.


\subsubsection{Alice sends a state}

The natural generalization of this protocol is to replace the Bell
states by arbitrary pure states generated by Alice
\cite{LC97,May97}. When these have the same restriction on Bob's
side, they are purifications of the same state, and hence
connected by a partial isometry on Alice's side, which serves as a
sneak flip operation. A crucial step is now to go away from
perfect concealment ($\varepsilon=0$ in Eq.~(\ref{eq:setup02})),
which seems to have been considered first in \cite{May97}. In this
case one has to use a continuity result for purifications, i.e.,
that nearby states have nearby purifications. In other words, one
needs an estimate \cite{NC00} of Uhlmann's fidelity (which
measures the distance between purifying vectors), and the trace
norm.


\subsubsection{Classical communication}

Classical communication occurs naturally in cryptographic
protocols, so it needs to be included in the analysis. In contrast
to some of our predecessors, who choose a purely quantum
description from the outset, we treat classical information
explicitly throughout. In particular, classical information in the
Lo-Chau-Mayers approach is treated quantum-mechanically and sent
over noiseless quantum channels, while our description explicitly
allows information transfer over classical channels, and thus
provides a natural setting to include purely classical protocols
in the analysis.

Cheating becomes harder for Alice if the protocol requires some
exchange of classical information, for she no longer has full
control over the purification spaces of the two commitment states.
Roughly speaking, unitaries which introduce superpositions of
states, which belong to different classical values already sent to
Bob, are forbidden. In the formalism introduced above this means
that Alice has to find a cheating unitary for every classical
communication history $x$.

Mayers' heuristic paper \cite{May97} has some provisions for this
case, by sending classical values to a special quantum repository
in the environment, and effectively coherentifying all classical
information. In contrast, in this text the classical
communication flow is treated explicitly, and in fact emerges
naturally as a framework for the description of the protocol. This
approach should also prove helpful in the analysis of other
cryptographic tasks.


\subsubsection{Bob supplies the paper}
    \label{sec:paper}

The protocols so far were characterized by the property that Bob
really had no strategic choices to make during the commitment
phase. Hence the state at the end of the commitment phase, written
in our scheme as $\rho_c(a,b)$, really does not depend on Bob's
strategy $b$. So Alice only has to connect the purifications of
two states which are explicitly known to her. Clearly, her task of
finding a clever sneak flip becomes harder if there is a proper
dependence on $b$. Lo-Chau-Mayers restrict their analysis to those
protocols in which Bob follows a specified `honest' strategy
$b_{\star}$, which is assumed to be publicly known in accordance
with Kerckhoffs' principle. In these cases, Alice knows how to
cheat, and the no-go result immediately applies.

As explained in the Introduction, we do not require that Bob
follows such a publicly known standard strategy. Alice then indeed
has to find a sneak flip working for all of Bob's admissible
strategies $b$. The easiest such protocol begins with Bob sending
a system to Alice, in some state known only to him (in
\cite{Yue05} this is called an {\it anonymous} state). The honest
strategies require Alice to encode the bit by using this system in
some way and then returning a committing system to Bob.
Effectively Alice now chooses not a {\it state} but a {\it
channel} to encode her commitment. The purification idea and
Uhlmann fidelity estimate no longer work for this, so these
protocols are not covered by Lo-Chau-Mayers. Instead, the
purification construction has to be generalized to the Stinespring
representation of channels, and an appropriate continuity result
has to be shown. This will be done in Section~\ref{sec:proof}.


\subsubsection{A decoherence monster in Bob's lab}

That the idea of states supplied by Bob may introduce interesting
new aspects is demonstrated by a scenario which is not a bit
commitment protocol in the sense of this section, because it makes
additional assumptions about things happening in Bob's lab:
Suppose that after Bob has sent some quantum state to Alice, a
Decoherence Monster (such as the cleaning service) enters his lab,
and all quantum information is destroyed. Only his classical
records survive. That is, he still knows what preparation he made,
but cannot use the entangled records he made during the
preparation. Now suppose that Alice and Bob can rely on this
happening. Then they can design a bit commitment protocol that
works. So, paradoxically, the monster strengthens Bob's position,
because it weakens the assumptions about his ability to break the
concealment. Hence one can make protocols which are binding in the
strong sense described above, but concealing only if we assume
that coherence in Bob's lab is indeed destroyed. We will analyze
this possibility in Section~\ref{sec:monster}.


\subsubsection{Alice can choose more strategies}

An apparent generalization would allow Alice to choose her honest
strategy $a_0$ at will from some set ${\sf A_0}$ of honest strategies, and $a_1$ from ${\sf A_1}$. The idea is that now some $a_0\in{\sf A_0}$ might
well be distinguishable from some $a_1\in{\sf A_1}$ for Bob.
Concealment under such circumstances means that Bob, on seeing
data compatible with some $a_0$ during the commitment or holding
phase can never be sure that they do not come from a certain
$a_1$. In other words, for every $a_0\in{\sf A_0}$ there must be
an $\varepsilon$-equivalent strategy $a_1\in{\sf A_1}$. But then,
according to our result, Alice might develop a sneak flip attack
on the basis of these two protocols alone.


\subsubsection{More communication in the holding phase}
    \label{sec:holding}

In Section~\ref{sec:english}, we excluded any communication in the
holding phase, and, apart from a single message from Alice to Bob,
also in the opening phase. There is, however, no problem to allow
such communication, and some protocols, like Kent's protocol using
relativistic signal speed constraints \cite{Ken99,Ken05}, require
a lot of communication in the holding phase.

Of course, protocols with no rounds at all in the holding phase
are directly covered by our definition. The only strategic
difference between holding and commitment phase is that Alice's
cheating strategies $a_0^\sharp$ and $a_1^\sharp$ are only
required to coincide during the commitment phase. She might start
cheating with different tricks for $0$ and $1$ during the holding
phase.

Clearly, declaring the holding phase a part of the commitment
phase only weakens Alice's cheating possibility.  However, she
does not need these extra options anyway: a sneak flip attack at
the end of the holding phase is always possible, as we show.


\subsubsection{Aborts and resets}

Often in cryptography one considers protocols which allow the
parties to call an ``abort''. We can distinguish two kinds of
abort: when a {\it constructive abort}, or {\it reset} occurs, the
protocol is started anew, whereas at a {\it full abort} the whole
protocol is terminated as unsuccessful.

Both kinds of aborts are covered in our scheme, but they would be
typical of different phases. Resets are quite natural in the
commitment phase. For example, Bob might make a test measurement
on some message he receives, and refuses to continue if there is a
slight deviation from what is expected from Alice playing honest.
A reasonable requirement at this point is that the probability for
reaching a commitment after some number of rounds with an honest
Alice is positive. Then allowing even more retrials one could
bring the probability for reaching commitment close to one, and
allow some arbitrary choice in the remaining cases, i.e., if the
allotted total number of rounds is exhausted without a commitment.
In this way one would get a protocol satisfying our finiteness
condition, while retaining the potential value of resets for a
commitment protocol. Strictly speaking, resets can only occur
during the commitment phase, since we have demanded a partitioning
of each protocol run into three successive phases (without
relapses into earlier phases). However, the holding phase can be
essentially united with the commitment phase (see
Sec.~\ref{sec:holding}). Hence we can effectively also cover
constructive aborts during the holding phase.

In the opening phase we can consider {\em full}, or destructive
aborts. This is a move right to an endpoint of the communication
tree, labelled accordingly. Clearly this possibility weakens Bob's
discrimination powers, and makes it much easier to cheat for
Alice. In particular, each sneak flip attack becomes successful.
Therefore, the abort possibility does not seem to present any
interesting strategic options for quantum bit commitment. The
proof in Sec.~\ref{sec:proof} shows that this is indeed the case.


\subsubsection{Concatenated protocols}

Sometimes one considers settings in which a variety of different cryptographic protocols are run in parallel or in succession, usually with dependent inputs. Obtaining bounds on the security of concatenated protocols in terms of the security parameters of their component parts is often far from straightforward, and a subject of ongoing research even in classical cryptography \cite{Can00}. However, in this work we are chiefly concerned with impossibility results, which easily transfer to concatenated protocols: Running a finite number of (possibly different) bit commitment protocols in parallel or succession, and assuming that those protocols all fall into the framework described in this chapter, the concatenated protocol is again a quantum bit commitment protocol, with suitably enlarged Hilbert and classical messenger spaces and possibly a larger number of rounds. Since the latter protocol is covered by our impossibility result, concatenating finitely many insecure bit commitment protocols cannot help to establish secure bit commitment.

The formulation of two party protocols that we describe in Sec.~\ref{sec:formal} is by no means limited to quantum bit commitment, and hence could also be used to model larger cryptographic environments, of which quantum bit commitment might be a subroutine. In Fig.~\ref{fig:tree}, such a protocol would appear as a subtree. Concealment and bindingness would have to be guaranteed for the entire tree, and hence by restriction for the subtree. Thus, no two-party cryptographic protocol covered by the framework described in Sec.~\ref{sec:formal} can contain a secure bit commitment protocol.

The composability analysis is of course much more involved for {\em secure} protocols. The security proof we provide in Sec.~\ref{sec:monster} for the decoherence monster protocol in general only applies to the protocol as a stand-alone object. If this protocol is then used as a subroutine in a larger and complicated cryptographic context, the security analysis will usually have to be tailored to the specific protocol. Fortunately, at least the cb-norm estimates we use in the proof of Th.~\ref{theo:monster} are stabilized distance measures, and hence well-behaved under concatenation (cf. App.~C).

\section{Proof}
    \label{sec:proof}

In the exposition of the task of bit commitment and the admissible
protocols we have tried not to restrict generality by simplifying
assumptions, in order not to weaken the scope of the no-go
theorem. This leads to a rather wild class of strategies to be
considered: arbitrarily many rounds of communication of varying
length, infinite-dimensional local lab Hilbert spaces, and all
that. Clearly, in the course of the proof we want to get rid of
this generality. The main idea for simplifications is that
obviously inferior methods of analysis for Bob, or inferior
cheating methods for Alice need not be considered. We therefore
begin with an explanation of what it means that one strategy is
``obviously inferior'', or {\it weaker} than another (see
Sec.~\ref{sec:compareStrat}).

The first application of this idea is the process of {\it
purification}, by which a general strategy is turned into another
one, which avoids all measurements not demanded by the
communication interface, and turns all decohering operations into
coherent information transfer to ancillas. Stinespring's dilation
theorem guarantees that this can always be done. We explain in Sec.~\ref{sec:purification} how the purifications
result in {\it locally coherent strategies}, which will be crucial
for Alice's cheat later on, and have been a part of all no-go
results.

Once a player has chosen a locally coherent strategy, it is
possible to reduce the lab spaces considerably. For example, if a
strategy requires the choice of a mixed state, this state may have
an infinite-dimensional support Hilbert space. Its purification,
however, is a single vector, so up to a unitary transformation,
which can be absorbed into subsequent operations, it suffices to
take a one dimensional Hilbert space. We show that this works for
operations as well: for every locally coherent strategy there is a
stronger one (in the sense of Sec.~\ref{sec:compareStrat}), using
only finite-dimensional Hilbert spaces, with a universal dimension
bound depending only on the dimension of quantum messages
exchanged so far and the trusted ressources shared initially. In particular, an infinite-dimensional lab space
will not give more power to Bob. This will be shown in
Sec.~\ref{sec:boundedH}, and leads to the consequence that
effectively (up to any desired level of accuracy) we need only
consider a finite number of strategies for Bob.

The next step is in some sense a dual of purification:
purification means that we can avoid measurements during a
protocol, deferring all such operations to the final measurement.
Similarly, we can move the acts of decision making during the
protocol to the very beginning, by introducing  a {\it strategy
register} (see Sec.~\ref{sec:register}), which is described in the
Hilbert space $\ell^2(S)$, for some finite set $S$ of strategies.
The choice of a strategy is then expressed by preparing some
initial state of the strategy register, and then letting
controlled unitaries transcribe this information into suitable
operations at all later rounds. Let us denote by $b_\sigma$ Bob's
strategy of installing the strategy register mechanism, and
preparing the initial state $\sigma$ for that register. The state
$\rho_c(a,b_\sigma)$ at commitment time then depends linearly on
$\sigma$, and after specifying the trusted initial state $\rho_0$ and tracing out Alice's lab, we find a channel
$\Gamma^{B}(a)$ depending on Alice's strategy $a$, such that
\begin{eqnarray}\label{gammA}
    \Gamma^{B}(a)&:&\bigoplus_{x\in X_c} \cB_x\to\bop{\ell^2(S)} \\
    {\rm tr} \; \sigma \; \Gamma^{B}(a)(B)
    &=& {\rm tr} \; \rho^B_c(a,b_\sigma) \; B
\end{eqnarray}
for all $B \in \oplus_{x} \cB_x$. This channel now summarizes
everything that Bob can possibly learn about Alice's strategy by
choosing his own strategy and making a measurement in his lab
after the commitment. In a simple, purely Kerckhoffian scenario
the analogous object is just the state at commitment time, since
one does not allow Bob a choice of different legitimate
strategies. However, in our more general framework we do need to
consider the dependence on $\sigma$, and correspondingly cheats
which work uniformly well for all $\sigma$.

As an instructive special case, we next suppose that the protocol
is {\it perfectly} concealing, which is expressed by
$\Gamma^{B}(a_0)=\Gamma^{B}(a_1)$. We show in
Sec.~\ref{sec:perfect} that Alice then has a perfect cheat. Its
existence is guaranteed by the uniqueness clause in the
Stinespring dilation theorem. From this prototype of Alice's cheat
one can see how an approximate cheat in response to approximate
concealment $\Gamma^{B}(a_0)\approx\Gamma^{B}(a_1)$ should work.

In the next section we look more carefully into the kind of
approximation $\Gamma^{B}(a_0)\approx\Gamma^{B}(a_1)$ sufficient
to draw the desired conclusion. It turns out that we need to
consider a special attempt of concealment breaking for Bob, namely
keeping an entangled record of the strategy register and making a
joint measurement on the rest of his system and this ``backup
copy'' after commitment. Clearly, this is a legitimate attempt in
our framework, and hence must already be implicit in the
strategies controlled by the strategy register. However, making
this scheme explicit provides the right kind of norm (cb-norm) on
channels so that a small
$||\Gamma^{B}(a_0)-\Gamma^{B}(a_1)||_{cb}$ guarantees the
existence of an approximately ideal cheat. The technical result
guaranteeing this is a new continuity theorem \cite{KSW06} for the
Stinespring dilation construction, which we review in
Sec.~\ref{sec:nogo}.


\subsection{Comparing the Strength of Strategies}
    \label{sec:compareStrat}

Consider two strategies $a$ and $a'$ of Alice. We will say that
$a'$ is {\it stronger} than $a$, if whatever Alice can achieve by
strategy $a$ she can also achieve by $a'$. More explicitly, we
require that there exists a suitable {\it revert} operation
$R_x:\cA_{x}(a)\to\cA_{x}(a')$ bringing Alice back to strategy $a$
at whatever node $x$ she so chooses (observe the direction of
arrows due to the Heisenberg picture). That she actually comes
back to $a$ is guaranteed inductively, i.e., we require that
\begin{align}
    \label{revert01}
        R_xT_x(a)&=T_x(a')\bigoplus_{m\in
        M_x}(R_{xm} \otimes\id_{\cM_m^x})\\
        & \qquad \qquad \quad \text{at Alice's nodes and} \nonumber\\
    \label{revert02}
        R_{xm}&= R_{x}\otimes\id_{\cM_m^x}\\
        & \qquad \qquad \quad \text{at Bob's nodes.}\nonumber
\end{align}
Tracing this all the way back to the root of the tree we get, for
any of Bob's protocols $b$, and for any stage of the protocol, in
particular for the commitment stage $X_c$,
\begin{equation}\label{revertstate}
   {\rm tr} \; \rho_c(a,b)\Bigl(\bigoplus_{x}F_x\otimes G_x\Bigr)
    = {\rm tr } \; \rho_c(a',b)\Bigl(\bigoplus_{x}R_x(F_x)\otimes
    G_x\Bigr).
\end{equation}
Taking $F_x=\idty_x$ in Eq.~(\ref{revertstate}) (corresponding to
the partial trace over Alice's lab space in the Schr\"odinger
picture), we see that Bob's subsystems are completely unaffected,
i.e., Bob will never be able to tell the difference between $a$
and $a'$. The strategic significance of passing to a stronger
strategy is different for Alice and for Bob.

{\it For Bob} a stronger $b'$ is just another strategy to be
considered in the concealing condition and in the condition for a
successful cheat. Since Bob does not loose any discriminating
power in playing coherent, Alice (and we) might as well assume
that he is always using the strongest strategy available. This
simplifies the analysis, as we will see in more detail below.

{\it For an honest Alice} there is no option. Whatever the honest
strategies $a_0$ and $a_1$ specify, she has to follow. However,
since Bob will never know the difference, it is easy to check from
the definitions of concealing and binding in Sec.~\ref{sec:formal}
that whenever $(a_0,a_1)$ is a bit commitment protocol with
security parameters $\varepsilon$ and $\delta$, then so is any
pair of stronger strategies $(a'_0,a'_1)$, with the same
parameters. Hence we could assume for the sake of an impossibility
proof that Alice's honest strategies are strengthened in some way.
However, there is hardly an advantage in that assumption, and we
will not do so.

{\it For a cheating Alice}, using all the power of her infinitely
well equipped lab, and hence using the strongest available
strategies is clearly the best choice. Indeed, this will be the
only difference between the honest and the cheating strategies
during the commitment phase: these consists of playing until
commitment a particular strengthening of an honest strategy,
namely the local purification discussed in the next subsection.


\subsection{Local Purification}
    \label{sec:purification}

Intuitively, maintaining coherence during quantum operations is
more demanding than allowing thermal noise and other sources of
decoherence to have their way. Therefore, doing only those
measurements  needed for satisfying the communication interface
rules, but avoiding all other decoherence should lead to a
stronger protocol in the sense of Sec.~\ref{sec:compareStrat}.

The simplified ``locally coherent'' strategies are more easily
expressed in terms of operators acting on Hilbert spaces than by
superoperators acting on algebras. Therefore we need a notation
for the message Hilbert spaces as well, i.e., we set
$\cM_m^x=\bop{\kk_m^x}$, where $\dim\kk_m^x=d(x,m)$ is the
dimension parameter from the description of the communication tree
in Sec.~\ref{sec:tree}.

\begin{define}
    \label{def:loccohere}
    {\bf (Locally Coherent Strategy)}\\
    We call a strategy $a$ of Alice {\em locally coherent} iff for
    all communication nodes $x$ we have $\cA_x(a)=\bop{\hh_x(a)}$ and, at all
    of Alice's nodes, the quantum channel
    $T_x(a):\bigoplus_m\cA_{xm}(a)\otimes\cM_m^x\to\cA_x(a)$ from
    Eq.~(\ref{eq:elementary01a}) is given by operators
    \begin{equation}\label{isostep01}
    V_{x,m}(a):\hh_x(a)\to
        \hh_{xm}(a)\otimes\kk_m^x
    \end{equation}
    such that
    \begin{equation}\label{isostep}
        T_x(a)\Bigl(\bigoplus_{m}A_m\otimes Y_m\Bigr)
        =\sum_{m} V_{x,m}(a)^*\bigl(A_m\otimes Y_m\bigr)V_{x,m}(a)
    \end{equation}
    for all
    $A_m\in\bop{\hh_{xm}(a)}$ and $Y_m\in\bop{\kk^x_{m}}$.
\end{define}

The point here is that each summand in this $T_x(a)$ is {\it
pure}, i.e., given by a single Kraus operator $V_{x,m}(a)$. This
is equivalent to the property that the $m^{\rm th}$ term in this
sum cannot be decomposed into a non-trivial sum of other
completely positive maps, which would in turn correspond to the
extraction of further classical information. Using a non-pure map
in a strategy would therefore mean to exercise less than the
maximal control allowed by quantum theory. Note that $m$ is in
general a random outcome, but Alice can make it deterministic by
choosing her strategy $a$ corresponding to
$V_{x,m}(a)=\delta_{m,m_0}V_x$, with an isometry $V_x$.

There is a canonical way to convert any strategy into a locally
coherent one, which is provided by the basic structure theorem for
completely positive maps. We state it in a form appropriate for
the finite-dimensional case which is needed here. We refer to
Paulsen's text \cite{Pau02} for further details and the proof.

\begin{propo}
    \label{stinespring}
    {\bf(Stinespring Dilation)}\\
        Let $\cA$ be a finite dimensional C*-algebra, $\hh$ a Hilbert space, and
        $T:\cA\to\bop\hh$ a completely positive map. Then there is another
        Hilbert space $\kk$, a *-representation $\pi:\cA\to\bop\kk$, and a
        bounded operator $V:\hh\to\kk$ such that, for all $A\in\cA$,
        \begin{equation}
            \label{eq:purification01}
                T(A)=V^*\pi(A)V.
        \end{equation}
        If $(\kk_0,\pi_0,V_0)$ and $(\kk_1,\pi_1,V_1)$ are two such
        representations, there is a partial isometry $U:\kk_0\to\kk_1$
        such that
        \begin{eqnarray}\label{uconnect}
        UV_0&=&V_1,\\
        U^*V_1&=&V_0 \qquad \qquad\mbox{and}\\
        U\pi_0(A)&=&\pi_1(A)U
        \end{eqnarray}
        for all $A\in\cA$.
\end{propo}

We will use this proposition several times, but ignore the
uniqueness statement for the moment. Then we can iteratively
generate a locally coherent protocol $\purify a$ from $a$,
together with the required revert operations showing that $\purify
a$ is indeed stronger than $a$. Suppose the space $\hh_x(\purify
a)$ and the revert channel $R_x:\cA_x(a)\to\bop{\hh_{x}(\purify a)}=\cA_x(\purify a)$ has already been defined along with these objects for all earlier nodes. We need to extend this definition to all successor nodes
$xm$. If the node $x$ belongs to Bob, there is nothing to do since
Eq.~(\ref{revert02}) explicitly defines $R_{xm}$. At Alice's
nodes, we apply the Stinespring Theorem to the composition
\begin{equation}
    \label{eq:purification02}
        R_xT_x(a):\bigoplus_{m\in M_x}\cA_{xm}(a)\otimes\cM_m^x \
       \to\bop{\hh_x(\purify a)}.
\end{equation}
The dilation theorem then provides us with a representation
$\pi_x$ of $\bigoplus_{m\in M_x}\cA_{xm}(a)\otimes\cM_m^x$ on some
Hilbert space $\kk_x$ and an isometry $V_x \mathpunct :
\hh_{x}(\purify a) \to \kk_{x}$. Now the projections $P_m$ in
$\bigoplus_{m\in M_x}\cA_{xm}(a)\otimes\cM_m^x$ which correspond
to the direct sum decomposition over $m$ are mapped by $\pi_x$ to
projections on $\kk_x$, so we get a decomposition into orthogonal
subspaces $\kk_{x}=\bigoplus_m\pi_{x}(P_m)\kk_{x}$. Since the
$P_m$ commute with all other elements of the algebra, the
projections $\pi_{x}(P_m)$ commute with all $\pi_{x}(A)$, and
$A\mapsto\pi_{x}(P_m)\pi_{x}(A)$ becomes a representation on
$\pi_{x}(P_m)\kk_{x}$. This representation can be restricted to
the message algebra $\cM_m^x$, and since the representation of a
full matrix algebra is unique up to multiplicity (and up to
unitary equivalence indicated by ``$\cong$'' in the equations
below), we can split the subspace $\pi_{x}(P_m)\kk_{x}$ into a
tensor product:
\begin{eqnarray}
    \label{splitdilate}
        \pi_{x}(P_m)\kk_{x}&\cong&\hh_{xm}
        (\purify a)\otimes\kk_m^x,\\
        \pi_{x}(\idty\otimes X) \, \pi_{x}(P_m)
        &\cong&\idty\otimes X,\\
        \pi_{x}(A\otimes\idty) \, \pi_{x}(P_m)
        &\cong&\pi_{xm}(A)\otimes\idty.
\end{eqnarray}
At the last line we have used that all $\pi_{x}(A\otimes\idty)$
commute with all $\pi_{x}(\idty\otimes X)\cong(\idty\otimes X)$,
so must be of the form $A'\otimes\idty$ for some $A'=\pi_{xm}(A)$.
We have already indicated in the notation that the space
$\hh_{xm}(\purify a)$ arising in this construction will be chosen
as Alice's lab Hilbert space for the coherent strategy $\purify
a$. The revert operation will simply be
$R_{xm}=\pi_{xm}:\cA_{xm}(a)\to\bop{\hh_{xm}(\purify a)}$ and,
finally, the isometries of the pure strategy will be
\begin{eqnarray}
    \label{eq:purification03}
            V_{x,m}(\purify a)\cong\pi_{x}(P_m)V_x(a) \mathpunct :
            \quad \hh_x(\purify a) & \to & \pi_{x}(P_m)\kk_{x}
            \\ & \cong & \hh_{xm}(\purify a)\otimes\kk_m^x.
            \nonumber
\end{eqnarray}
Then Eq.~(\ref{revert01}) holds by virtue of the Stinespring
representation, and we have shown that $\purify a$ is indeed
stronger than $a$.

To summarize: for every strategy $a$ there is a stronger locally
coherent strategy $\purify a$. Moreover, the corresponding revert
operation can be chosen to be a representation for all $x$. Of course, the same construction holds for Bob's nodes.

In the sequel we will assume from now on that Bob uses coherent strategies, since this does not constrain his power to resolve Alice's actions at
any stage. As we will show in the proof of Th.~\ref{theo:nogo}, Alice's cheat consists in playing suitable purified strategies, too. By means of Eq.~(\ref{revertstate}), purification on Alice's side will give Bob no clue whatsoever about her cheating attempt.


\subsection{Bounding Local Hilbert Space Dimensions}
    \label{sec:boundedH}

It is a crucial point in the definition of concealment that no
limitations are imposed on Bob's capabilities. In particular, he
could choose to use arbitrarily large local lab Hilbert spaces. In
principle, this makes scanning all of Bob's strategies for
checking $\varepsilon$-concealment an infinite task. However, the
purification construction takes care of this aspect as well, and
we will show that without loss of discrimination power Bob can fix
the dimension of his lab spaces uniformly over all his strategies.

The Stinespring construction respects finite dimensionality.
Usually one takes a ``minimal'' dilation, which means that the
vectors $\pi(A)V\phi$ with $A\in\cA$ and $\phi\in\hh$ are dense in
$\kk$. Hence $\dim\kk\leq\dim\cA\cdot\dim\hh$. However, since this
bound still contains the algebra $\cA$, which is part of the
strategy whose purification generates the locally coherent
protocol, and which is not a priori bounded, this argument does
not suffice to derive a uniform dimension bound on local lab
spaces.

The desired bound can be constructed by looking directly at the
definition of locally coherent strategies. Here the growth of
Bob's lab space is given by the two operations
\begin{eqnarray}
   V_{x,m}(b)&:&\hh_x(b)\to \kk_m^x \otimes \hh_{xm}(b)\\
             && \qquad \qquad \qquad \text{at Bob's nodes and} \nonumber\\
   \hh_{xm}(b)&=&\kk_m^x\otimes\hh_{x}(b) \\
             && \qquad \qquad \qquad \text{at Alice's nodes.} \nonumber
\end{eqnarray}
Given the dimensions of $\hh_x(b)$ and $\kk_m^x$, the first line
per se does not imply a bound on the dimension of $\hh_{xm}(b)$.
However, the range of $V_{x,m}$ has known finite dimension, so
most of these dimensions will never be used. More precisely, we
can find a subspace $\hh'_{xm}(b)\subset\hh_{xm}(b)$ such
that\begin{equation}
    \label{sufficientdim}
        V_{x,m}(b) \, \Bigl(\hh_x(b)\Bigr)
        \subset \kk_{m}^{x} \otimes \hh'_{xm}(b).
\end{equation}
Indeed, we can take ${\hh'_{xm}(b)}$ as the span of all vectors
$\phi_{\alpha,j}$ appearing in the expansion $V_{x,m} (b) \,
\phi_\alpha=\sum_j \psi_j \otimes \phi_{\alpha,j}$, where
$\{\psi_j\}\subset\kk_m^x$ and $\{\phi_\alpha\}\subset\hh_x(b)$
are orthonormal bases. Hence
\begin{equation}\label{dimbound}
    \dim {\hh'_{xm}(b)}\leq\dim\hh_{x}(b)\,\dim\kk_m^x.
\end{equation}

We can now apply this idea inductively, i.e., with a previously
constructed $\hh_x'(b)\subset\hh_x(b)$ on the left hand side of
Eq.~(\ref{sufficientdim}). Note that at Alice's nodes there is
nothing to choose, and the dimension bound Eq.~(\ref{dimbound})
holds with equality anyhow. At the root we have
$\dim\hh_0(b)=\dim\hh_0'(b)=:d_{0}^{B} \in \N$ for all strategies,
the dimension of Bob's initial state space.

Hence we have a new strategy, using the same isometries
$V_{x,m}(b)$ as $b$, but with domains and ranges restricted to a
subspace $\hh_x(b')\equiv\hh'_x(b)\subset\hh_x(b)$ for all $b$. We
show now that $b'$ is stronger than $b$. The required revert
operation is implemented by the subspace embedding $j_x \mathpunct
: \hh_x(b')\to\hh_x(b)$, as $R_x(B)=j_x^*Bj_x$ and, due to
Eq.~(\ref{sufficientdim}), the operators $V_{x,m}$ for the new
strategies are connected by
\begin{equation}
    V_{xm}(b)\, j_x=(\idty\otimes j_{xm}) \, V_{xm}(b') \mathpunct
    : \hh_x(b')\to \kk_m^x\otimes\hh_{xm}(b),
\end{equation}
where $j_{xm}$ is the embedding of $\hh_{xm}'(b)$ into $\hh_{xm}(b)$.
Eq.~(\ref{revert01}) then follows by combining this with
Eq.~(\ref{isostep}) in a version adapted to Bob's pure strategies.
An intuitive description of this revert operation in the
Schr\"odinger picture is to ask Bob to consider his density
operator on $\hh_x(b')$ as a density operator on the larger space
$\hh_x(b)$, by setting it equal to zero on the orthogonal
complement.

It is perhaps paradoxical that in this case the strategy using less
resources is stronger. But in fact, they are just equally strong.
The revert operation in the opposite direction is
$S_x:\bop{\hh_x(b')}\to\bop{\hh_x(b)}$, with
\begin{equation}\label{expandrevert}
    S_x(B)=j_xBj_x^*+ \rho_x(B)(\idty-j_xj_x^*),
\end{equation}
where $\rho_x$ is an arbitrary state on $\bop{\hh_x(b')}$. The
second term is added to satisfy the channel normalization
$S_x(\idty)=\idty$. Since $j_x^*j_x=\idty$, we have $R_xS_x=\id$.
The revert operation in this case is thus the projection on the
subspace $\hh_{x}(b') \subset \hh_{x}(b)$.

Taking together the reduction operation, and, possibly an
expansion as described (adding some extra dimensions on which all
states vanish), we can convert any strategy $b$ to another one,
for which the dimension bound Eq.~(\ref{dimbound}) holds with
equality, at both Bob's and Alice's nodes. But then we can
identify all the spaces $\hh_x(b')$ with a fixed space of
appropriate dimension, say $\hh_{x}^{B}$.

Applying the same construction to Alice's operations, we find a
strategy-independent Hilbert space $\hh_{x}^{A}$. In particular,
we will henceforth assume $\hh_{x}(\purify a_0) = \hh_{x}(\purify
a_1) = \hh_{x}^{A}$ at all nodes $x$ for Alice's locally coherent
strategies $\purify a_i$. This will simplify the discussion of
Alice's cheating strategy in Secs.~\ref{sec:perfect} and
\ref{sec:nogo} below.

We summarize this section in the following Proposition, which we
formulate for Bob's strategies. It holds equally for Alice's
strategies, too.
\begin{propo}
    \label{propo:bounded}
        {\bf (Dimension Bound)}\\
        Let $\hh_{x}^B$ denote a family of Hilbert spaces
        with dimensions satisfying
        \begin{eqnarray}
            \label{eq:boundedH01}
                \dim {\hh_{xm}^B} & = &
                \dim\hh_{x}^B\,\dim\kk_m^x\\
                \mbox{and\ } \qquad
                \dim\hh_0^B & = & d_{0}^{B} \in \N
        \end{eqnarray}
        for all nodes $x$. Then for every locally coherent strategy $b$
        of Bob there is an equally strong locally coherent strategy $b'$
        with $\hh_x(b')=\hh_x^B$ for all $x$.
\end{propo}
The entire strategy dependence is now contained in the choice of
the operators $V_{x,m}(b')$.
\begin{coro}
    \label{cor:approx}
        In the definitions of $\varepsilon$-concealing and
        $\delta$-cheating strategy, we may restrict the quantifier over
        all of Bob's strategies to locally coherent strategies with a
        strategy-independent lab Hilbert space $\hh^B_x$.\\
        For every $\xi>0$ there is a finite set $S$ of such strategies
        approximating all of Bob's discriminating procedures to within
        $\xi$. That is, for any strategy $b$ of Bob we can find $b'\in S$
        such that for all of Alice's strategies $a$:
        \begin{equation}
            \label{eq:boundedH02}
                \norm{\rho_c(a,b)-\rho_c(a,b')}_1 \leq\xi.
        \end{equation}
\end{coro}
The proof of Corollary~\ref{cor:approx} is obvious from the
dimension bound, and the observation that the set of bounded
operators between Hilbert spaces of fixed finite dimension is
compact in the norm topology.


\subsection{Bob's Strategy Register}
    \label{sec:register}

The next simplification we would like to introduce will
significantly reduce the complexity of the many-round scenario.
The basic idea is to replace all of Bob's choices by a single
choice he makes at the beginning by preparing a suitable initial
state. His later choices will then be taken over by a sequence of
``quantum controlled operations''. This reorganization of Bob's
choices requires the expansion of the lab space by an additional
register, to hold the control information. It is perhaps
worthwhile to emphasize that this strategy register serves merely
as a technical tool in the no-go proof.

We will choose a finite approximation $S$ to Bob's strategy space
in the sense of Corollary~\ref{cor:approx}, with a very small
value of $\xi$, which will be taken to zero at the end. The
strategy register will be described by the Hilbert space
$\ell^2(S)$, the complex valued functions on $S$, with the usual
scalar product. In other words, we have one basis vector $\ket b$
for each strategy $b\in S$. Then we set
\begin{eqnarray}
    \label{eq:register01}
        \widetilde\hh^B_x&=&\hh^B_x\otimes\ell^2(S)\\
        \widetilde V_{x,m}&:&\widetilde\hh^B_x\to
        \widetilde\hh^B_{xm}\otimes\kk_m^x\\
        \widetilde V_{x,m}&=&\sum_{b\in S} V_{x,m}(b)\otimes \kettbra b
\end{eqnarray}

Observe that $\tilde V_{x,m}$ is now independent of Bob's strategy
(it depends on $S$). However, Bob still has a choice to make,
namely the choice of the initial state for the strategy register.
If he wants to play strategy $b$, he will set it to $\kettbra b$
and then let the pre-programmed controls take over.

The construction also opens up the rather interesting possibility
for Bob to play strategies in superposition, simply by initially
preparing a superposition of the basis states $\ket b$. For this
case it is helpful to bear in mind that the ``control'' by
``controlled unitary operations'' is not a one way affair. As soon
as Bob prepares superpositions, the strategy register is in
general affected by the interaction, so by ``measuring the
strategy'' after a while, Bob could pick up some clues about
Alice's actions. This is required by basic laws of quantum
mechanics, because the controlled-unitary operation creates
entanglement.

Let us consider the overall effect of the protocol up to
commitment, with the trusted shared initial state $\rho_0$ considered fixed, Bob choosing an arbitrary initial state $\sigma
\in \bops{\ell^{2}(S)}$ (possibly mixed) for the strategy
register, and Alice playing strategy $a$. At commitment, the
observable algebra is now $\bigoplus_{x\in X_c}\cA_x (a)
\otimes\bop{\widetilde\hh_x^B}$. The state obtained on this
algebra depends linearly on the initial state $\sigma$, and being
implemented by a series of completely positive transformations,
this dependence is given by a quantum channel $\chan(a)$. In the
Heisenberg picture we thus have
\begin{equation}\label{bigchan}
    \chan(a):\bigoplus_{x\in X_c}\cA_x(a)\otimes\bop{\widetilde\hh_x^B}
       \to \bop{\ell^{2}(S)}.
\end{equation}
The restriction of the final state to Bob's side is what decides
his chances of distinguishing different strategies of Alice. These
restrictions are given by the reduced channel
$\chan^B(a):\bigoplus_{x\in X_c}
   \bop{\widetilde\hh_x^B}\to\bop{\ell^2(S)}$, given by
\begin{equation}
    \label{eq:register02}
        \chan^B(a)\Bigl(\bigoplus_{x \in X_c}B_x\Bigr)
        =\chan(a)\Bigl(\bigoplus_{x \in X_c}\idty_{\cA_x(a)}\otimes B_x\Bigr).
\end{equation}
The concealment condition requires that
$\chan^B(a_0)\approx\chan^B(a_1)$. The aim of the impossibility
proof is to conclude from this the existence of a good cheating
strategy for Alice. For this conclusion it turns out to be crucial
how the approximate equality of these channels is expressed
quantitatively. We defer this discussion to Sec.~\ref{sec:record},
and treat first the case $\chan^B(a_0)=\chan^B(a_1)$, which
requires only the Stinespring dilation theorem, and shows more
clearly what properties we need to establish in the approximate
case.


\subsection{The Case of Perfect Concealment}
    \label{sec:perfect}

In the sequel Bob is always understood to take advantage of his
strategy register and pre-programmed controls, as described in
Sec.~\ref{sec:register}. So we will henceforth drop the tilde on
Bob's Hilbert spaces $\widetilde{\hh}_{x}^{B}$ to streamline the
presentation.

For the case of perfect concealment, suppose that
$\chan^B(a_0)=\chan^B(a_1)$, and that Alice is preparing to cheat.
She will then play the local purification $\purify a_i$ ($i=0,1$)
of one of the honest strategies until commitment time. Note that
both Alice's and Bob's strategies are assumed to be locally
coherent in the sense of Sec.~\ref{sec:purification}, with Hilbert
space dimensions independent of their respective strategies as
explained in Sec.~\ref{sec:boundedH}. The concatenated channel
$\chan(\purify a_i) \mathpunct : \oplus_{x} \bop{\hh_{x}^{A}}
\otimes \bop{\hh_{x}^{B}} \rightarrow \bop{\ell^{2}(S)}$ is then
likewise pure, and is hence given by operators
$V_{i,x}:\ell^2(S)\to\hh_{x}^{A} \otimes \hh_{x}^{B}$ as
\begin{equation}
    \label{eq:perfect01}
        \begin{split}
            \chan(\purify a_i)\Bigl(\bigoplus_{x\in X_c}(A_x\otimes B_x)\Bigr)
            & =\sum_{x \in X_c} V_{i,x}^*(A_x\otimes B_x)V_{i,x}\\
            & =V_{i}^*\Bigl(\,\bigoplus_{x\in X_c}(A_x\otimes
            B_x)\Bigr)\,V_{i}.
        \end{split}
\end{equation}
For the last step we have combined all the $V_{i,x}$ into a single
operator $V_{i}:\ell^2(S)\to\kk := \bigoplus_x\hh_{x}^{A} \otimes
\hh_x^B$, and the direct sum refers to the direct sum
decomposition of the underlying Hilbert space $\kk$. Note that
this Hilbert space carries a representation $\pi$ of Bob's
observable algebra $\bigoplus_x\bop{\hh_x^B}$ at commitment time,
simply by setting $\pi(\bigoplus_xB_x)=\bigoplus_x\idty_x^A\otimes
B_x$. Hence $(\kk,\pi,V_i)$ is a dilation of the channel
$\chan^B(\purify a_i)$ in the sense of Prop.~\ref{stinespring}.

But now, by assumption $\chan^B(\purify
a_0)=\chan^B(a_0)=\chan^B(a_1)=\chan^B(\purify a_1)$. Hence we get
two dilations of the same channel, which must be connected by a
unitary operator $U \in \bk$ as in Prop.~\ref{stinespring}.
Essentially, this $U$ will be Alice's cheat operation. What we
have to show is that she can execute this operation on the system
under her control, given the classical information $x$.

The condition $U\pi(Y)=\pi(Y)U$, applied to a projection $Y=P_x$
of one of the summands implies that $U$ can be broken into blocks,
$U\pi(P_x)=\pi(P_x)U \in \bop{\hh_{x}^{A} \otimes \hh_{x}^{B}}$.
The intertwining relation for $\pi(B_x)$ allows us to conclude
that this operator is of the form $U_x\otimes\idty_x^B$, with a
unitary operator $U_x \in \bop{\hh_{x}^{A}}$. Clearly, $U_x$ is an
operator between possible lab spaces of Alice, depending only on
publicly available information $x\in X_c$. This will be Alice's
cheat channel. Setting
\begin{eqnarray}\label{cheatchan}
    C_x \mathpunct : \bop{\hh_x(\purify a_1)}\to\bop{\hh_x(\purify a_0)}
    \qquad
    C_x(A)=U_x^*AU_x,
\end{eqnarray}
we immediately conclude from $UV_0=V_1$ that
\begin{equation}\label{cheateq}
    \chan(\purify a_0)(\bigoplus_xC_x\otimes\id_x^B)=\chan(\purify
    a_1).
\end{equation}

Let us summarize Alice's perfect cheat. She will play the
purification $\purify a_0$ of the honest strategy $a_0$ until
commitment time. If at that time she decides to go for the bit
value 0, she will just apply the revert operation from the
purification construction. After that nobody can tell the
difference between her actions and the honest $a_0$, not even with
full access to both labs. On the other hand, if she wants to
choose bit value 1, she will apply the cheat channel $C_x$. We see
from Eq.~(\ref{cheateq}) that afterwards nobody will be able to
tell the difference between her actions and $\purify a_1$.
Finally, she will apply the revert operation from $\purify a_1$ to
$a_1$, hiding all her tracks. Note that the revert operation by
construction works at any step: indeed Alice can cheat at any
time, since the protocol must be concealing for all steps in order
to be concealing at the commitment stage.


\subsection{Bob's Entangled Strategy Record}
    \label{sec:record}

In the previous section we have seen how Stinespring's theorem
allows Alice to find a perfect cheat in a perfectly concealing bit
commitment protocol. The continuity theorem presented in
Section~\ref{sec:nogo} below shows that the same cheating strategy
still works for Alice with high probability under more realistic
conditions --- when only approximate concealment is guaranteed,
$\chan^{B}(a_0)\approx\chan^{B}(a_1)$. The result crucially
depends on the way in which the distance between these two
channels is evaluated: Bob can test the condition
$\chan^{B}(a_0)\approx\chan^{B}(a_1)$ by preparing a state
$\sigma$ for the strategy register $\ell^2(S)$, and making a
measurement on the system $\hh^B_x$ he receives back from Alice.
This includes both the possibility to superpose his original
strategies $\ket b$, and the possibility to mix such strategies in
the sense of game theory. However, this still does not exhaust his
options: he can {\it keep an entangled record of his strategy}.
This would be pointless for just classical mixtures of his basic
strategies $\ket b$. In that case all his density operators would
commute with the ``strategy observable'', and he could extract the
initial strategy by a von Neumann measurement from the state at
any later step. However, if he also uses superpositions of
strategies, the controlled unitaries may properly ``change'' the
strategy. It therefore makes sense to keep a record, i.e., to not
only use a mixed initial state, which would correspond to a mixed
strategy in the sense of von Neumann's game theory, but to use an
entangled pure state on $\ell^2(S)\otimes\ell^2(S')$, with some
reference system $S'$. It turns out that one can always choose $S'
\cong S$ (cf. Prop.~8.11 in Paulsen's text \cite{Pau02}). While
the first copy in this tensor product is used as before to drive
the conditional strategy operators $V_x$, the second is the record
and is completely left out of the dynamics. In other words, Bob
not only uses a von Neumann mixed strategy, but the purification
of this mixture. Concealment will then have to be guaranteed
against his joint measurements on $\hh_B^{x}\otimes\ell^2(S')$.

We will see in Sec.~\ref{sec:monster} that this procedure in
general does increase Bob's resolution for the difference of
channels. Of course, if the initial selection of strategies $S$ is
large enough, an approximation of this quantum randomized strategy
will already be contained in $S$, and the gain may be negligible.
Mathematically, the introduction of randomized strategies
corresponds to using a different norm: to guarantee concealment in the sense of Def.~\ref{def:concealing}, Alice will have to make
sure that $\norm{\big ( \chan^{B}(a_0) - \chan^{B}(a_1) \big )
\otimes \id_{\nu}} \leq \varepsilon$ if $\nu$-dimensional
bystander systems are taken into account, for all $\nu\in\N$. As
explained in the Appendix, this just means that these two channels
need to be indistinguishable in cb-norm, $\cbnorm{\chan^{B}(a_0) -
\chan^{B}(a_1)} \leq \varepsilon$ for some small $\varepsilon >0$.


\subsection{The Full Impossibility Proof}
    \label{sec:nogo}

The full impossibility proof goes beyond the case of perfect
concealment discussed in Sec.~\ref{sec:perfect}. It shows that
Alice can still cheat if the bit commitment protocol is only
approximately concealing, and provides explicit
dimension-independent bounds on Alice's probability to pass Bob's
tests undetected:
\begin{theo}
    \label{theo:nogo}
        {\bf (No-Go Theorem)}\\
        Any $\varepsilon$-concealing bit commitment protocol in
        the sense of Sec.~\ref{sec:formal} allows Alice to find a
        $2 \, \sqrt{\varepsilon}$-cheating strategy.
\end{theo}
These bounds coincide with those obtained by Spekkens and Rudolph
\cite{SR01} in the Kerckoffian framework. Our proof shows that
they still hold if Bob no longer sticks to a publicly known
strategy. This is a significant improvement over Cheung's
dimension-dependent estimates \cite{Che06}, which do not suffice
to rule out bit commitment protocols with large systems.

The full no-go proof is based on a continuity result for
Stinespring's dilation theorem, which we cite here from
\cite{KSW06}. It states that two quantum channels $\chan_{0}^{B}$,
$\chan_{1}^{B}$ are close in cb-norm iff there exist corresponding
Stinespring isometries $V_{0}$, $V_{1}$ which are close in
operator norm. This generalizes the uniqueness clause in
Stinespring's theorem to cases in which two quantum channels
differ by a finite amount, and hence is precisely the type of
result we need to rule out approximately concealing bit commitment
protocols.
\begin{propo}
    \label{propo:continuity}
        {\bf (Continuity Theorem)}\\
        Let $\hh^{}$ and $\hh^{B}$ be finite-dimensional Hilbert spaces,
        and suppose that
        \begin{equation}
            \label{eq:nogo01}
                \chan_{0}^{B}, \chan_{1}^{B} \mathpunct : \quad
                \mathcal{B} (\hh^{B}) \rightarrow \bh
        \end{equation}
        are quantum channels with Stinespring isometries $V_{0},
        V_{1} \mathpunct : \hh^{} \rightarrow \hh^{A} \otimes
        \hh^{B}$ and a common dilation space $\hh^{A}$ such that $\dim \hh^{A} \geq 2 \, \dim \hh_{} \, \dim \hh^{B}$. We then
        have:
        \begin{equation}
            \label{eq:nogo02}
                \begin{split}
                    \inf_{U} \norm{(U_{} \otimes \idty_{B}) V_{0} -
                    V_{1}}^{2} & \leq \cbnorm{\chan_{0}^{B} -
                    \chan_{1}^{B}}\\
                    & \leq 2 \, \inf_{U} \norm{(U_{}
                    \otimes \idty_{B}) V_{0} - V_{1}},
                \end{split}
        \end{equation}
        where the minimization is over all unitary $U \in
        \mathcal{B} (\hh^{A})$.
\end{propo}
We refer to \cite{KSW06} for a proof of
Prop.~\ref{propo:continuity} and further applications of the
continuity theorem. In this form the result applies to quantum
channels whose common domain is a full matrix algebra, while in
our case the domain algebra of the commitment channels
$\chan^{B}_{i} \equiv \chan^{B}(\purify a_i)$ is the direct sum
$\oplus_x \mathcal{B} (\hh_{x}^{B})$. Again we have dropped the
tilde from Bob's Hilbert spaces in an attempt to streamline the
presentation. In order to apply the continuity theorem to our
setting, we extend the channels $\chan_{i} \equiv \chan (\purify
a_i) \mathpunct : \oplus_{x} \bop{\hh_{x}^{A}} \otimes
\bop{\hh_{x}^{B}} \rightarrow \bh$ to channels $\hat\Gamma_0,
\hat\Gamma_1\mathpunct: \bop{\hh^A\otimes\hh^B} \to \bop\kk$,
where we have introduced the shortcuts $\hh := \ell^{2}(S)$,
$\hh^A :=\oplus_x\hh^A_x$ and $\hh^B := \oplus_x\hh^B_x$. Note
that the tensor product $\hh^A\otimes\hh^B$ has the direct sum
decomposition $\oplus_{xy}\hh_x^A\otimes\hh_y^B$, and that
$\oplus_x\bop{\hh_x^A\otimes\hh_x^B}$ is the subalgebra in
$\bop{\hh^A\otimes\hh^B}$ which consists of those operators that
are supported on the diagonal subspace
$\oplus_{x}\hh_x^A\otimes\hh_x^B$. For direct sum channels
$\Gamma_{i} ( \oplus_{x} A_x \otimes B_x ) = \sum_{x} V_{i,x}^{*}
(A_x \otimes B_x) V_{i,x}^{}$ as in Eq.~(\ref{eq:perfect01}), the
extensions $\hat\Gamma_i=\hat V_i^*(\cdot)\hat V_i^{}$ have
Stinespring isometries $\hat V_0,\hat
V_1\mathpunct:\hh^{}\to\hh^A\otimes\hh^B=\oplus_{xy}\hh_x^A\otimes\hh_y^B$
given by
\begin{equation}
    \label{eq:nogo03}
        \hat V_i\psi:=\bigoplus_{xy}\delta_{xy}V_{i,x}\psi \; .
\end{equation}
In the sequel we assume that the dilation spaces are chosen sufficiently large such that the dimension bound in Prop.~\ref{propo:continuity} is met. The restrictions of $\hat{\Gamma}_{i}^{}$ to Bob's output system
$\hh^{B}$ will be denoted by $\hat{\Gamma}^{B}_{i}$. We then have
$\hat\Gamma_i^B=\Gamma_i^B\circ P$, where the cp-map
\begin{equation}
    \label{eq:nogo03a}
            P \mathpunct : \bop{\hh^B} \to \oplus_{x} \,
            \bop{\hh^B_x} \qquad
            P (B) = \oplus_x \, P_x B P_x
\end{equation}
is composed of the projections $P_{x}$ in $\hh^B$ onto $\hh_x^B$.
Since
\begin{equation}
    \label{eq:nogo04}
        \cbnorm{\hat\Gamma_0^B-\hat\Gamma_1^B}=
        \cbnorm{(\Gamma_0^B-\Gamma_1^B)\circ
        P}\leq  \cbnorm{\Gamma_0^B-\Gamma_1^B},
\end{equation}
we may now apply the left half of the continuity estimate
Eq.~(\ref{eq:nogo02}) to the extended quantum channels
$\hat{\Gamma}_{i}^{B}$ to conclude that
\begin{equation}
    \label{eq:nogo05}
        \inf_{U} \, \norm{ (U\otimes\idty_{B})
        \hat V_{0} - \hat V_{1}}^2 \leq
        \cbnorm{\hat\Gamma^B_{0}-\hat\Gamma^B_{1}}
        \leq \cbnorm{\hat\Gamma_{0}-\hat\Gamma_{1}}.
\end{equation}
The minimization at this point is with respect to all unitary
$U\in\bop{\hh^A}$, which can be given the block decomposition
\begin{equation}
    \label{eq:nogo06}
        U\psi=\bigoplus_x \sum_y U_{xy}\psi_y
\end{equation}
with operators $U_{xy}:\hh^A_y\to\hh^A_x$. It turns out that the
minimization in Eq.~(\ref{eq:nogo05}) can always be restricted to
unitary operators whose off-diagonal blocks vanish. To see this,
note that the left hand side of Eq.~(\ref{eq:nogo05}) can be
rewritten as
\begin{equation}
    \label{eq:nogo07}
        \begin{split}
            \inf_{U} & \, \norm{ (U\otimes\idty_{B}) \hat V_{0} - \hat
            V_{1}}^2 \\
            & = \inf_U\, \sup_{\varrho} {\rm tr} \; \varrho \big ( \hat
            V_0^*(U^*\otimes\idty_{B})-\hat V_1^* \big ) \, \big (
            (U\otimes\idty_{B}) \hat V_{0} - \hat V_{1} \big )\\
            & = \inf_U \, \sup_{\varrho}  \Bigl ( 2 - 2 \, {\rm Re} \, {\rm tr}
            \, \varrho \, \hat V_1^*(U\otimes \idty_B)\hat V_0 \Bigr)
        \end{split}
\end{equation}
where the supremum is taken over all states $\varrho \in \bhstar$.
From the definition of the isometries $\hat{V}_{i}$ in
Eq.~(\ref{eq:nogo03}) above it is straightforward to verify that
\begin{equation}
    \label{eq:nogo08}
        \hat V_1^*(U\otimes \idty_B)\hat V_0=\sum_x
        V_{1,x}^*(U_{xx}\otimes\idty_x)V_{0,x}
\end{equation}
in Eq.~(\ref{eq:nogo07}). Therefore, the minimization procedure on
the left hand side of Eq.~(\ref{eq:nogo05}) is not affected by the
off-diagonal blocks $\{U_{xy},x\not=y\}$, which implies that the
infimum is attained at a unitary operator that is a direct sum of
unitaries, $U=\oplus_x U_x\in\oplus_x \bop{\hh^A_x}$. On the other
hand, the cb-norm difference $\cbnorm{\Gamma_{0}^{B} -
\Gamma_{1}^{B}}$ is easily seen to be upper bounded by $2 \,
\norm{(U \otimes \idty_{B}) V_{0} - V_{1}}$ for any unitary
operator $U = \oplus_{x} U_{x}$.

In summary, we have shown that the continuity theorem extends to
direct sum channels with a unitary $U$ that respects the
direct-sum decomposition:
\begin{propo}
    \label{propo:sumcontinuity}
        {\bf (Continuity Theorem for Direct Sum Channels)}\\
        Let $\hh$ be a finite-dimensional Hilbert space, and let
        $\{\hh_{x}^{B} \}_{x \in X}$ and $\{\hh_{x}^{A} \}_{x \in
        X}$ be collections of finite-dimensional Hilbert spaces.
        Suppose that $V_{0}, V_{1} \mathpunct : \hh \rightarrow \oplus_{x}
        \hh_{x}^{A} \otimes \hh_{x}^{B}$ are Stinespring
        isometries for the quantum channels $\Gamma_{1},
        \Gamma_{2} \mathpunct : \oplus_{x} \bop{\hh_{x}^{A} \otimes
        \hh_{x}^{B}} \rightarrow \bh$ such that
        \begin{equation}
            \label{eq:nogo09}
                \begin{split}
                    \Gamma_{i} \Bigl(\bigoplus_{x}(A_x\otimes B_x)\Bigr)
                    & =\sum_xV_{i,x}^*(A_x\otimes B_x)V_{i,x}\\
                    & =V_{i}^*\Bigl(\,\bigoplus_x(A_x\otimes
                    B_x)\Bigr)\,V_{i}
                \end{split}
        \end{equation}
        and $\dim \hh_{x}^{A} \geq 2 \, \dim \hh_{x}^{B} \, \dim \hh$ for all $x \in X$.
        Let $\Gamma_{i}^{B} \mathpunct : \oplus_{x} \bop{\hh_{x}^{B}}
        \rightarrow \bh$ be the local restrictions given by
        $\Gamma_{i}^{B} (\oplus_{x} B_x ) := V_{i}^{*} (\oplus_{x}
        \idty_{x}^{A} \otimes B_{x}) V_{i}^{}$. We then have:
        \begin{equation}
            \label{eq:nogo10}
                \begin{split}
                    \inf_{U} \norm{(U_{} \otimes \idty_{B}) V_{0} -
                    V_{1}}^{2} & \leq \cbnorm{\chan_{0}^{B} -
                    \chan_{1}^{B}} \\
                    & \leq 2 \, \inf_{U} \norm{(U_{}
                    \otimes \idty_{B}) V_{0} - V_{1}},
                \end{split}
        \end{equation}
        where the minimization is over all unitary operators $U =
        \oplus_{x} U_{x} \in \oplus_{x} \, \bop{\hh_{x}^{A}}$.
\end{propo}
The proof of the no-go theorem now immediately follows from
Prop.~\ref{propo:sumcontinuity}.\\
\\
{\bf Proof of Th.~\ref{theo:nogo}:} Alice will play the
purification $\purify a_0$ of the honest strategy $a_0$ until
commitment time. If at that time she decides to go for the bit
value $0$, she will just apply the revert operation $R$ from the
purification construction, as described in
Sec~\ref{sec:purification}. It is then no longer possible to tell
the difference between her actions and the honest $a_0$, not even
with full access to both labs. On the other hand, if she wants to
choose bit value $1$, she will apply the cheat channel $C_x
\mathpunct : \bop{\hh_{x}(\purify a_1)} \rightarrow
\bop{\hh_{x}(\purify a_0)}$ given by $C_x (A) := U_{x}^{*} A
U_{x}^{}$, where $U : = \oplus_{x} U_{x} \in \oplus_{x} \,
\bop{\hh_{x}^{A}}$ is the unitary operator that attains the
infimum in Eq.~(\ref{eq:nogo10}) above. In the purification construction detailed in Sec.~\ref{sec:purification} we have for simplicity assumed minimal dilation spaces. Yet in order to apply the sneak flip operation, Alice may possibly need to double her local lab space $\oplus_{x} \,
\bop{\hh_{x}^{A}}$ to satisfy the dimension bound in Prop.~\ref{propo:sumcontinuity}. However, this can always be postponed right until before the cheat, only requires an additional (sufficiently large) ancilla system, and hence does not constrain Alice's options.

Given an $\varepsilon$-concealing bit commitment protocol with local
channels $\chan^{B} (a_i)$ in the sense of Def.~\ref{def:concealing}, we conclude from our discussion in Sec.~\ref{sec:record} that $\cbnorm{\chan^{B}(a_0) - \chan^{B}(a_1)} \leq \varepsilon$. Hence, the continuity estimate
implies that
\begin{equation}
\label{eq:nogo11}
\begin{split}
& \hspace{-20pt}\cbnorm{\chan (\purify a_0) \big ( \bigoplus_x C_x \otimes
        \id_{x}^{B} \big )  -  \chan (\purify a_1)}
        \\
        &\leq 2\norm{\left( U\otimes \idty^B \right )
        V(\purify a_0)-V(\purify a_1)}
        \\
        &\leq 2\sqrt{\cbnorm{\chan^{B}(a_0) -
        \chan^{B}(a_1)}} \  \leq \ 2\sqrt{\varepsilon} \; ,
\end{split}
\end{equation}
where $V(\purify a_0), V(\purify a_1)$ are Stinespring isometries for $\chan (\purify a_0)$ and
$\chan (\purify a_1)$, respectively. Since the cb-norm difference cannot increase under quantum
channels, the same bound holds after Alice's revert operation $R$,
\begin{equation}
    \label{eq:nogo12}
        \cbnorm{\chan (\purify a_0) \big ( \bigoplus_x C_x \otimes
        \id_{x}^{B} \big ) \, R - \chan (a_1)} \leq 2
        \sqrt{\varepsilon}.
\end{equation}
Alice can then confidently announce the bit value $1$ in the
opening. The probability of her cheat being detected is upper
bounded by $2 \, \sqrt{\varepsilon}$. This concludes the proof of
the strengthened no-go theorem. $\blacksquare$


\section{QBC in Infinite Dimensions}
    \label{sec:energy}

In this Section we will relax the general finiteness condition
imposed in Secs.~\ref{sec:setup} and \ref{sec:proof} and show how
to extend the no-go proof to quantum bit commitment protocols in
which the dimension of the underlying Hilbert spaces
(Sec.~\ref{sec:cont}), the number of rounds
(Sec.~\ref{sec:rounds}), or the set of classical signals
(Sec.~\ref{sec:inftree}) are infinite.


\subsection{Continuous Variable Systems}
    \label{sec:cont}

We have so far restricted the discussion of the no-go theorem to
systems that can be described in finite-dimensional (albeit
arbitrarily large) Hilbert spaces. In this section we show that
the results can be easily extended to continuous variable systems
--- as long as the systems obey a global energy constraint of a reasonably generic
form. The total available energy for the protocol needs to be
finite but can otherwise be as high as desired, and yet secure
quantum bit commitment remains impossible. Purists might dismiss
this additional energy constraint on the basis that it restricts
the domain for the impossibility proof. Yet most physicists know
that infinite energy is seldom available. In fact, the continuity
theorem for Stinespring's dilation may be generalized to
completely positive maps between arbitrary $C^{*}$-algebras
\cite{KSW07}, and hence the no-go theorem applies to continuous
variable systems with unbounded energy, too. But these results are
somewhat beyond the scope of the present paper, so we assume a
uniform energy constraint to simplify the presentation.

To set the stage, assume that $\hh$ is a separable (but no longer
necessarily finite-dimensional) Hilbert space. As before, let
$\bhstar$ denote the Banach space of trace-class operators on
$\hh$, and $\mathcal{S} (\hh) \subset \bhstar$ the closed convex
set of states. We further assume that $H \mathpunct : \mathcal{D}
\rightarrow \hh$ is an unbounded self-adjoint (energy) operator
defined on a dense set $\mathcal{D} \subset \hh$. (From the
Hellinger-Toeplitz theorem (cf. Sec.~III.4 in \cite{RS80}) we know
that a symmetric unbounded operator cannot be defined on all of
$\hh$, so we always assume a dense subset $\mathcal{D}$.) For the
proof we assume that $H$ has discrete spectrum, that all of its
eigenvalues $h_{n}$ have finite multiplicity, and that
$\lim_{n\to\infty} h_{n} = \infty$. Under these conditions, the
set of states
\begin{equation}
    \label{eq:energy01}
        \mathcal{S}_{E} (\hh) := \big \{ \varrho \in \mathcal{S}
        (\hh) \; \mid \; {\rm tr} \, \varrho \, H \leq E \big \}
\end{equation}
can be shown to be compact for every $E \geq 0$ \cite{Hol03}. As
we assume this energy constraint to be global, we impose that it
is respected by the quantum operation $T_{*}$ that describes the
full bit commitment protocol: $T_{*}(\varrho) \in \mathcal{S}_{E}
(\hh)$ for all $\varrho \in \mathcal{S}_{E} (\hh)$.

Since the continuity theorem applies in this setting \cite{KSW06},
the proof presented in Sec.~\ref{sec:proof} goes trough unchanged.
There is also a simpler proof, which avoids the compactness
arguments and is based on a useful approximation result: any
infinite-dimensional system with energy constraints as in
Eq.~(\ref{eq:energy01}) can be approximated to arbitrary degree of
accuracy by a sufficiently large finite-dimensional system. This
allows to reduce any bit commitment protocol to its
finite-dimensional counterpart:
\begin{propo}
    \label{propo:reduction}
        Given an $\varepsilon$-concealing and $\delta$-binding
        quantum bit commitment protocol with a global energy
        constraint as in Eq.~(\ref{eq:energy01}). Then for any $\gamma > 0$
        there is a corresponding protocol on finite-dimensional
        Hilbert spaces with dimension $d = d(\gamma)$
        which is $(\varepsilon + \gamma)$-concealing
        and $(\delta + \gamma)$-binding.
\end{propo}
Since the latter protocol is unfeasible for sufficiently small
parameters $\varepsilon$, $\delta$ and $\gamma$, so is the former.

The finite-dimensional approximation needed for the proof of
Prop.~\ref{propo:reduction} relies on the following two lemmas:
\begin{lemma}
    \label{lemma:energy}
        Let $\gamma > 0$ and $\mathcal{S}_{E} (\hh)$ as in Eq.~(\ref{eq:energy01}).
        Then there exists a finite-dimensional projector
        $P_{\gamma}$ such that
        \begin{equation}
            \label{eq:energy02}
                {\rm tr} \, \varrho \, P_{\gamma} \geq 1 - \gamma
                \quad \forall \quad \varrho \in \mathcal{S}_{E} (\hh).
        \end{equation}
\end{lemma}
As a consequence, every system with energy constraints is
essentially supported on a finite-dimensional Hilbert space.
\begin{lemma}
    \label{lemma:channel}
        Let $\gamma > 0$ and $P_{\gamma}$ as in
        Lemma~\ref{lemma:energy}. Then for every quantum channel
        $T_{*} \mathpunct : \bhstar \rightarrow \bhstar$ which respects
        the energy constraint Eq.~(\ref{eq:energy01}) we have:
        \begin{equation}\label{eq:energy03}
\begin{split}
                \tracenorm{T_*(\varrho) - \frac{1}{{\rm tr} \,
                P_{\gamma} T_*(P_{\gamma} \varrho P_{\gamma})}
                P_{\gamma} \, & T_* (P_{\gamma} \varrho P_{\gamma}) \,
                P_{\gamma} }\\ &\leq 4 \, \sqrt{\gamma} + \frac{2
                \gamma}{1 - \gamma}
\end{split}
        \end{equation}
        for all $\varrho \in \mathcal{S}_{E} (\hh).$
\end{lemma}
The proof of Prop.~\ref{propo:reduction} is then straightforward:
Given the continuous-variable bit commitment protocol with energy
bound $E$ and security parameters $\varepsilon$ and $\delta$, we
construct its finite-dimensional companion by projecting on the
subspace $P_{\gamma} \hh$, with the finite-dimensional projector
$P_{\gamma}$ chosen as in Lemma~\ref{lemma:energy}. We know from
the discussion in Sec.~\ref{sec:proof} that both the concealment
and the bindingness condition can be expressed in terms of
appropriately chosen quantum channels $T_{*}$. By assumption,
these will respect the energy constraint. The approximation in
Lemma~\ref{lemma:channel} then guarantees that for sufficiently
small $\gamma$ the companion protocol has nearly identical
security parameters. Substituting $4 \, \sqrt{\gamma} + \frac{2
\gamma}{1 - \gamma} \mapsto \gamma$, this concludes the proof.

It remains to prove the approximation lemmas. The proof of
Lemma~\ref{lemma:energy} appears in \cite{Hol03}. We include
it here for completeness:\\
\\
{\bf Proof of Lemma~\ref{lemma:energy}:} Let the eigenvalues of
$H$ be arranged in increasing order: $h_{1} \leq h_{2} \leq h_{3}
\leq \ldots$, with eigenprojector $P_{n}$ corresponding to the
eigenvalue $h_{n}$. For $N \in \N$, we set $\hat{P}_{N} :=
\sum_{n=1}^{N} P_{n}$. We then have for all $\psi \in \hh$:
\begin{eqnarray}
    \label{eq:energy04}
            \brakket{\psi}{h_{N+1} \big ( \idty - \hat{P}_{N} \big
            )}{\psi}  &=&  \brakket{\psi}{h_{N+1}
            \sum_{n=N+1}^{\infty} P_{n}}{\psi} \nonumber \\
            & \leq & \brakket{\psi}{\sum_{n=N+1}^{\infty} h_{n}
            P_{n}}{\psi} \nonumber \\
            & \leq & \brakket{\psi}{H}{\psi},
\end{eqnarray}
implying that $h_{N+1}(\idty - \hat{P}_{N}) \leq H$ for all $N \in
\N$. We may then conclude that
\begin{equation}
    \label{eq:energy05}
        {\rm tr} \, \varrho \big (\idty - \hat{P}_{N} \big ) \leq
        \frac{1}{h_{N+1}} {\rm tr} \, \varrho \, H \leq
        \frac{E}{h_{N+1}}
\end{equation}
for all $\varrho \in \mathcal{S}_{E} (\hh)$. Since the sequence
$\{h_{N}\}_{N}$ diverges, the result follows by choosing
$P_{\gamma} :=
\hat{P}_{N_0}$ for some sufficiently large $N_0$. $\blacksquare$\\
\\
{\bf Proof of Lemma~\ref{lemma:channel}:} An application of the
triangle inequality shows that
\begin{eqnarray}
    \label{eq:energy06}
        \tracenorm{\varrho - P_{\gamma} \varrho P_{\gamma}} &\leq&
        \tracenorm{\varrho - P_{\gamma} \varrho} +
        \tracenorm{P_{\gamma} \varrho - P_{\gamma} \varrho
        P_{\gamma}} \nonumber \\
        &\leq& \tracenorm{(\idty - P_{\gamma}) \varrho} +
        \tracenorm{\varrho (\idty - P_{\gamma})}.
\end{eqnarray}
For $\varrho \in \mathcal{S}_{E} (\hh)$ we know from
Lemma~\ref{lemma:energy} that ${\rm tr} \, (\idty - P_{\gamma})
\varrho \leq \gamma$, and thus the two terms on the right of
Eq.~(\ref{eq:energy06}) may be bounded as follows:
\begin{eqnarray}
    \label{eq:energy07}
        \tracenorm{\big (\idty - P_{\gamma} \big ) \varrho} & = & {\rm
        tr} \, U \big (\idty - P_{\gamma} \big ) \varrho \nonumber \\
        &\leq& {\rm tr}^{\frac{1}{2}} \sqrt{\varrho} U^{} U^{*}
        \sqrt{\varrho} \; \; {\rm tr}^{\frac{1}{2}} \sqrt{\varrho}
        \big ( \idty - P_{\gamma} \big ) \sqrt{\varrho} \nonumber \\
        & \leq & \sqrt{\gamma},
\end{eqnarray}
where we have used the Cauchy-Schwarz inequality for the
Hilbert-Schmidt inner product, and $U$ denotes the polar isometry
of $(\idty - P_{\gamma}) \varrho$. Analogously, we have
$\tracenorm{\varrho (\idty - P_{\gamma})} \leq \sqrt{\gamma}$,
which together with Eqs.~(\ref{eq:energy06}) and
(\ref{eq:energy07}) implies that
\begin{equation}
    \label{eq:energy08}
        \tracenorm{\varrho - P_{\gamma} \varrho P_{\gamma}} \leq 2
        \, \sqrt{\gamma}.
\end{equation}
For all $\varrho \in \mathcal{S}_{E} (\hh)$, the renormalized
state $\frac{1}{{\rm tr} P_{\gamma} \varrho} P_{\gamma} \varrho
P_{\gamma}$ satisfies the estimate
\begin{equation}
    \label{eq:energy09}
        \frac{1}{{\rm tr} P_{\gamma} \varrho} P_{\gamma} \varrho
        P_{\gamma} - P_{\gamma} \varrho P_{\gamma} \leq
        \frac{\gamma}{1 - \gamma} P_{\gamma} \varrho
        P_{\gamma},
\end{equation}
which in combination with Eq.~(\ref{eq:energy08}) implies that
\begin{equation}
    \label{eq:energy10}
        \tracenorm{\varrho - \frac{1}{{\rm tr} P_{\gamma} \varrho}
        P_{\gamma} \varrho P_{\gamma}} \leq 2 \, \sqrt{\gamma} +
        \frac{\gamma}{1- \gamma}.
\end{equation}
Since the trace norm cannot increase under quantum operations
\cite{NC00}, the upper bound also holds for the norm difference
$\tracenorm{T_{*}(\varrho) - \frac{1}{{\rm tr} P_{\gamma} \varrho}
T_{*}(P_{\gamma} \varrho P_{\gamma})}$. As the quantum channel
$T_{*}$ is supposed to respect the energy constraint
Eq.~(\ref{eq:energy01}), an analogous chain of estimates for the
output states of the channel and yet another application of the
triangle inequality then yield the desired result. $\blacksquare$


\subsection{An Infinite Number of Rounds}
    \label{sec:rounds}

In this Section we will show how the no-go proof can be extended
to cover quantum bit commitment protocols with a possibly infinite
number of rounds --- as long as the expected number of total
rounds remains finite. Just as with the energy constraints
discussed in Sec.~\ref{sec:cont}, for any practical purpose this
additional assumption does not restrict the domain of the
impossibility proof.

We begin by explaining how the framework introduced in
Sec.~\ref{sec:setup} can be modified to easily accommodate bit
commitment protocols with an infinite number of rounds in each the
commitment, holding, and opening phase. As in the previous Section,
the impossibility proof will then follow from an approximation
argument.

As described in detail in Sec.~\ref{sec:formal}, each layer $X_t$
of the communication tree consists of a finite number of nodes.
Each node $x\in X_t$ is connected with nodes $xm\in X_{t+1}$ of
the following layer, corresponding to the classical message $m\in
M_x$. However, there is no longer a definite layer for which a
commitment or opening has been reached. Instead, there are now
infinitely many layers $X_t, t\in{\mathbb N}$. If Alice and Bob
choose a definite pair of strategies, they check by means of
suitable measurements how many rounds $t$ have been performed, and
whether they are willing to continue. The number of rounds then
naturally plays the role of a classical random variable.
Introducing the bundle of algebras
\begin{equation}
\underline{{\cal F}}\mathpunct : \; t \mapsto \underline{{\cal
F}}_t:=\bigoplus_{x\in X_t}\cA_x\otimes\cB_x \, ,
\end{equation}
the total system is now described by the algebra
$C(\underline{{\cal F}})$ of all bounded sections
\begin{equation}
F\mathpunct : t\mapsto F(t)\in \underline{{\cal F}}_t \; ,
\end{equation}
where the norm in $C(\underline{{\cal F}})$ is the standard
supremum norm given by $\|F\|=\sup_{t\in\N}\|F(t)\|$.

Alice's observable algebra, which we denote by
$C(\underline{\cA})$, is the subalgebra in $C(\underline{{\cal
F}})$ which consists of all bounded sections $A$ that assign to
every number of rounds $t$ an operator $A(t)$ belonging to Alice's
subsystem:
\begin{equation}
A:t\mapsto A(t)\in\underline{\cA}_t:=\bigoplus_{x\in
X_t}\cA_x\otimes\idty_{B,x} \, .
\end{equation}
The observable algebra $C(\underline{\cB})$ of Bob's system is
defined completely analogously. In our setup for protocols, each
strategy $a$ that Alice chooses is related to a channel
\begin{equation}
\Gamma(a):C(\underline{{\cal F}})\to\cB(\ell^2(S)) \, ,
\end{equation}
containing all her possible responses to strategies that Bob can
play by a suitable preparation of his strategy register
$\ell^2(S)$.

The channels $\Gamma(a)$ include naturally the necessary tests to
decide for each round whether to remain in the commitment phase or
to proceed with the holding or opening phase. The measurement of
the number of rounds corresponds to the embedding of the abelian
$C^*$-algebra of bounded functions on $\N$ -- denoted by $C(\N)$
--, which is obviously a subalgebra of $C(\underline{{\cal F}})$.
Let $\delta_t$ be the function in $C(\N)$ which takes the value
$\delta_t(t)=1$ and $\delta_t(s)=0$ if $s\not=t$, and let $\sigma$
be some state on $\cB(\ell^2(S))$ determining Bob's strategy.
Then, by definition, the quantity
\begin{equation}
P(a,\sigma|t):=\tr(\sigma \, \Gamma(a)(\delta_t))
\end{equation}
is the probability that the commitment phase has been reached at
round $t$, provided Alice plays $a$ and Bob plays $\sigma$.

As advertised above, we now impose the reasonable assumption that
whenever Alice plays honestly the expected number of rounds until
commitment is uniformly bounded in the choice of Bob's initial
state $\sigma$: denoting by $a_0$ and $a_1$ Alice's honest
strategies to commit either $0$ or $1$, respectively, there is a
finite constant $T \in \R$ such that
\begin{equation}
\sup_{\sigma}\sum_{t\in \N}P(a_i,\sigma|t) \ t  \leq T
\label{equ-bound}
\end{equation}
holds for $i=0,1$. The basic idea for the proof of the no-go
result is now to relate this bound to the energy bound of the
previous subsection, and hence approximate a protocol with
possibly infinitely many rounds by a protocol with ``a priori"
finitely many moves. As ``Hamiltonian", the
number-of-rounds-operator
\begin{equation}
H=\sum_{t\in\N} \delta_t
\end{equation}
is perfectly suited. In line with Eq.~(\ref{eq:energy01}), it is then enough to ensure that the
states $\Gamma(a_i)_*(\sigma)$ lie inside
$\mathcal{S}_{E} (L_2(\underline{\hh}))$ for an appropriately
chosen constant $E$, where $L_2(\underline{\hh})$ is the Hilbert
space of square integrable sections
\begin{equation}
\psi:t\mapsto\psi(t)\in\underline{\hh}_t:=\bigoplus_{x\in X_t}\hh^A_x\otimes\hh^B_x \; .
\end{equation}
Indeed, we conclude from Eq.~(\ref{equ-bound}) that for each initial state $\sigma$ the inequality
\begin{equation}
\tr(\Gamma(a_i)_*(\sigma)H)=\sum_{t\in\N}\tr(\sigma \Gamma(a_i)(\delta_t))\leq T
\end{equation}
is fulfilled for $i=0,1$. Hence, $\Gamma(a_i)_*(\sigma)\in\mathcal{S}_{T}(L_2(\underline{\hh}))$, and we immediately obtain the following corollary as consequence of Prop.~\ref{propo:reduction}:

\begin{coro}
    Suppose an $\epsilon$-concealing and $\delta$-binding protocol with infinitely many rounds such that the expected number of rounds until commitment is uniformly bounded for Alice playing honest as in Eq.~(\ref{equ-bound}). Then for any $\gamma > 0$
    there is a corresponding protocol on a priori finite number of rounds $N(\gamma)$ which is $(\varepsilon + \gamma)$-concealing
    and $(\delta + \gamma)$-binding.
\end{coro}

Thus, even protocols with an infinite number of rounds
do not admit unconditional secure bit commitment.


\subsection{A Continuous Communication Tree}
    \label{sec:inftree}

\begin{figure}[htb]
    \psfrag{A}{Alice}
    \psfrag{B}{Bob}
    \psfrag{c}{$X_{t+2}$}
    \psfrag{a}{$X_t$}
    \psfrag{b}{$X_{t+1}$}
    \psfrag{c}{$X_{t+2}$}
    \psfrag{p}{$\phi_t$}
    \psfrag{q}{$\phi_{t+1}$}
    \psfrag{j}{$x$}
    \psfrag{i}{$\phi_{t+1}(x)$}
    \psfrag{h}{$\phi_t\phi_{t+1}(x)$}
    \psfrag{t}{steps}
    \includegraphics[width=0.9\columnwidth]{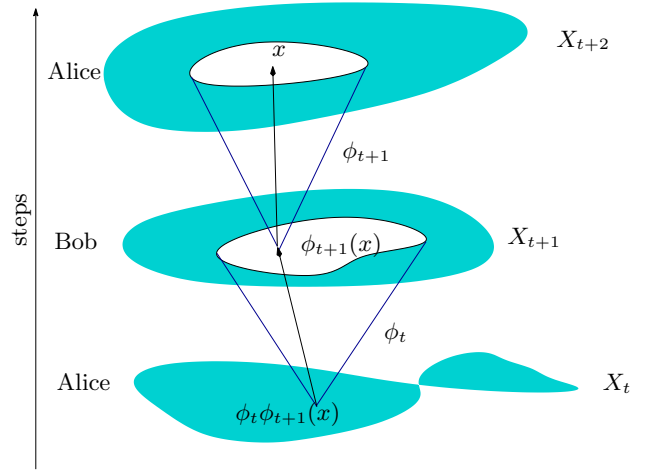}
    \caption{\label{fig:cont-tree}{\it Part of a path in a continuous communication tree. The layer $X_t$ belongs to Alice's move. She sends the message $(\phi_{t+1}(x),\phi_t\phi_{t+1}(x))$ to Bob. The next layer is Bob's turn, and consists in sending the message $(x,\phi_{t+1}(x))$ to Alice.}}
\end{figure}

So far we have assumed that our protocol is based on a communication tree with a discrete set of nodes and a finite number of options or messages. In this subsection we are going to relax this condition by allowing that the nodes as well as the options are taken from an {\em continuous set}. The set of time steps, however, is kept discrete. For simplicity, we restrict the discussion to protocols with a fixed number of rounds $c$ until commitment. How does the structure of a continuous communication tree look like? Each layer $X_t$ that is associated with the number of the time step $t$ is now taken to be a continuous compact manifold. The continuous version of connecting each ``node" of the layer $X_t$ to some nodes of the following slice $X_{t+1}$ is given by a continuous surjective map $\phi_t\mathpunct:X_{t+1}\to X_t$. The pre-image $\phi_t^{-1}(x)\in X_{t+1}$ of a point $x\in X_t$ corresponds to the nodes in $X_{t+1}$ that are connected to $x$. A pair $(x,\phi_t(x))$, $x\in X_{t+1}$, is regarded as a classical message that was sent from the node $\phi_t(x)\in X_t$ and was received at the node $x\in X_{t+1}$. Fig.~\ref{fig:cont-tree} shows a part of such a continuous communication tree. As with discrete communication trees, we associate to each message $(x,\phi_t(x))$ of the continuous tree a message system that is given by a full matrix algebra $\cM(x,\phi_t(x))=\cB(\kk_{(x,\phi_t(x))})$. Furthermore, to each point $x\in X_t$ we associate a finite dimensional full matrix algebra for Alice $\cA(x)=\cB(\hh_x^A)$, and likewise for Bob $\cB(x)=\cB(\hh_x^B)$. These algebras are defined recursively: if $t+1$ is Alice's turn, then her algebra $\cA(x)$ can be chosen arbitrarily. Bob's algebra is defined in terms of his previous choices of algebras and the message systems, as follows:
\begin{equation}
\cB(x):=\cM(x,\phi_t(x))\otimes\cB(\phi_t(x))
\end{equation}
for each $x\in X_{t+1}$. Since $X_t$ is continuous, we need to replace direct sums by bounded sections within a bundle of observable algebras. Introducing the bundle of algebras
\begin{equation}
\underline{\cF}_t:X_t\ni x\mapsto\underline{\cF}_t(x):=\cA(x)\otimes\cB(x) \, ,
\end{equation}
the total system at time step $t$ is described in terms of the $C^*$-algebra $C(\underline{\cF}_t)$ of bounded continuous sections in $\underline{\cF}_t$.  Alice's subsystem $C(\underline{\cA}_t)$ is determined by the constraint that for each $A\in C(\underline{\cA}_t)$ the value $A(x)$ is contained in the algebra $\cA(x)$. In other words, $A$ is a section in the sub-bundle $\underline{\cA}_t\mathpunct:x\mapsto\cA(x)$. Bob's system $C(\underline{\cB}_t)$ is defined in the same manner.

For every strategy $a$ played by Alice we hence obtain a corresponding channel $\Gamma(a): C(\underline{\cF}_c)\to\cB(\ell^2(S))$ modeling all the local operations and the entire exchange of messages up until commitment time $t=c$. As described in Sec.~\ref{sec:register}, Bob's strategies are programmed by the choice of the initial state $\sigma$ on $\cB(\ell^2(S))$. Just as before, these channels can be decomposed into a sequence of operations, each corresponding to a move made by Alice or Bob:
\begin{equation}
\label{eq:chain}
\Gamma(a)=T_{(a,1)}\circ T_{(a,2)}\circ\cdots \circ T_{(a,c)} \; .
\end{equation}
We may assume that Alice and Bob play locally coherent strategies. Then all the channels
\begin{equation}
T_{(a,t)}:C(\underline{\cF}_{t+1}) \to C(\underline{\cF}_t)
\end{equation}
in the decomposition Eq.~(\ref{eq:chain}) are pure. Depending on whose turn it is, the channels $T_{(a,t)}$ need to respect Alice's and Bob's
subsystems: if $t+1$ is Alice's move, then $T_{(a,t)}$ maps Alice's subsystem $C(\underline{\cA}_{t+1})$ into Alice's subsystem of the previous step  $C(\underline{\cA}_t)$, whereas Bob's system $C(X_{t+1},\underline{\cB})$ remains unaffected. Consequently,
\begin{eqnarray}
        T_{(a,t)}(A)(x) & \in & \cA(x) \qquad {\rm and}\\
        T_{(a,t)}(B\circ \phi_t)(x) &=& B(x)
\end{eqnarray}
for all $x\in X_t$ and for all $B\in C(\underline{\cB}_t)$. Note that for each section $B\in C(\underline{\cB}_{t})$ corresponding to Bob's system the section
\begin{equation}
B\circ\phi_t:y\mapsto B(\phi_t(y))\in\cB(\phi_t(y))\subset \cB(y)
\end{equation}
also belongs to Bob's system  $C(\underline{\cB}_{t+1})$ at the preceding step.

In complete analogy to the concealing condition for the discrete tree, for a protocol with a continuous communication tree to be $\epsilon$-concealing we require that
\begin{equation}
    \label{eq:con}
        \cbnorm{\Gamma^{B}(a_0)-\Gamma^{B}(a_1)}\leq \epsilon \, ,
\end{equation}
where $\Gamma^{B}(a)$ is the restriction of $\Gamma(a)$ to Bob's subsystem $C(\underline{\cB}_{c})$. The range and domain algebras of the channels $\Gamma^{B}(a_i)$ are no longer finite dimensional matrix algebras, nor do they admit a straightforward approximation in terms of finite dimensional systems. Nevertheless, the concealing condition Eq.~(\ref{eq:con}) implies that Alice may find a $2 \, \sqrt{\varepsilon}-$cheating strategy, as before. The result follows from a generalization of the continuity theorem for Stinespring's representation theorem to general $C^{*}-$algebras \cite{KSW07}.


\section{Protocols Relying on Decoherence}
    \label{sec:decoherence}

In this Section we will demonstrate how trusted decoherence
in Alice's lab (Sec.~\ref{sec:shredder}), Bob's lab
(Sec.~\ref{sec:monster}), or in the transmission line
(Sec.~\ref{sec:transmission}) may be employed to design
secure and fair bit commitment protocols.


\subsection{The Trusted Coherence Shredder}
    \label{sec:shredder}

A trusted third party makes perfect bit commitment a trivial task:
Alice may submit the bit to an incorruptible notary public, who
will store the bit in his vault throughout the holding phase, and
later pass it on to Bob on Alice's notice. In this scenario, the
notary public will have to be paid for the long-term safe storage
of the bit. Clearly, Alice and Bob would get away with much lower
fees, if the notary's presence were only required once, and only
as a witness, without even having to store a file about the event.
Such a possibility is offered by quantum mechanics.

The basic idea is that the notary is present in Alice's lab until
the end of the commitment phase, and sees to it that Alice plays
honest. If the honest protocols were locally coherent, even that
would be no help, since we have seen that Alice could carry out
her cheating transform later, in the holding phase. However, if
the honest protocols $(a_0,a_1)$ involve some measurement or other
decoherence, the notary overseeing these actions can make a
difference. He could prevent a later cheat by taking some part of
the system with him and destroying it. In our example below it is
even sufficient for him to just watch Alice make a measurement
and, if he so chooses, to forget about the result straight away.
The protocol is perfectly concealing, and is as binding as
desired, if a dimension parameter $d$ is chosen large enough.

The setting requires a $d$-dimensional Hilbert space, and two
mutually unbiased orthonormal bases $\{\ket{e_j}\}_{j}$,
$\{\ket{f_k}\}_{k}$, which means that
$\braket{e_j}{e_k}=\braket{f_j}{f_k}=\delta_{jk}$, and
$|\braket{e_j}{f_k}|^2=1/d$, for all $j,k=1,\ldots,d$. While the
maximum number of mutually unbiased bases in a Hilbert space of
given dimension $d$ is the subject of ongoing research, here we
only need two such bases, which are always easily constructed:
starting from any given orthonormal basis $\{\ket{e_j}\}_{j}$, we
may choose $\{\ket{f_k}\}_{k}$ as the Fourier-transformed basis,
\begin{equation}
    \label{eq:locdec01}
        \ket{f_{k}} := \frac{1}{\sqrt{d}} \sum_{j=1}^{d}
        e^{\frac{2 \pi i}{d} j k} \ket{e_{j}}.
\end{equation}

The protocol begins by Alice sending Bob a half of the maximally
entangled state
\begin{equation}
    \label{eq:locdec02}
        \ket{\Omega}=\frac1{\sqrt d}\sum_j\ket{e_j}\otimes\ket{e_j}
         =\frac1{\sqrt d}\sum_j\ket{f_j}\otimes\ket{\overline{f_j}},
\end{equation}
where $\ket{\overline{f_j}}$ denotes the complex conjugate of
$\ket{f_j}$ with respect to the basis $\ket{e_j}$. Then, if she
wants to commit the bit value ``$0$'', she makes a von Neumann
measurement in the basis $\ket{e_j}$, and records the result.
Similarly, to commit a ``$1$'', she makes a measurement in the
basis $\ket{f_j}$. Thus, if she plays honest, as vouched for by
the notary public, she will have no quantum system left in her
lab, only the classical information about the bit value, and her
measurement result. This is the information she sends to Bob at
the opening. To verify, he will make a measurement in the basis
$\ket{e_j}$, if Alice claims to have submitted ``$0$'', and in the
basis $\ket{\overline{f_j}}$ otherwise, finding the same result as
Alice with probability $1$.

The protocol is perfectly concealing, since in either case Bob
gets a system in the chaotic state $\rho_B=\frac1d\idty$. It is
also binding, because whatever false bit value and measurement
result Alice claims, Bob will confirm this only with probability
$1/d$, i.e., practically never, if $d$ is large.

This is essentially the bit commitment protocol originally
proposed by Bennett and Brassard in 1984 \cite{BB84}. Alice's EPR
attack does not work in our scenario, since the notary public will
not permit her to delay the measurements until after the
commitment phase. There is also a variant of this protocol, in
which the measurement is not actually carried out. In that case
Alice prepares one of the mixed states
\begin{align}
    \label{eq:locdec03}
        \rho_0&=\frac1d\sum_j\kettbra{e_j\otimes e_j}, \\
    \label{eq:locdec04}
        \rho_1&=\frac1d\sum_j\kettbra{f_j\otimes \overline{f_j}},
\end{align}
for committing ``$0$'' or ``$1$'', respectively. Now the notary
watching her will see to it that she actually prepares these mixed
states, and not their purifications. For verification Bob uses the
support projections $P_{0,1}=d\cdot\rho_{0,1}$.

Once again, the protocol is perfectly concealing. Let us analyze
Alice's cheating options, after she prepared $\rho_0$, with the
trusted notary watching and then leaving. If she wants to change
her commitment to ``$1$'', she can only employ some local channel
$T\otimes\id$ and hope to pass Bob's test with the projection
$P_1$. The probability for this is
\begin{eqnarray}
    \label{eq:locdec05}
        \tr & \rho_0 & \! \! \! (T\otimes\id)(P_1) \nonumber \\
     &=&\frac1d\sum_{k,j=1}^{d}\bra{e_j , e_j}
         (T\otimes\id)(\kettbra{f_k , \overline{f_k}})
         \ket{e_j , e_j}\nonumber\\
     &=&\frac1d\sum_{k,j=1}^{d}\abs{\braket{e_j}{\overline{f_k}}}^2\
         \bra{e_j} T(\kettbra{f_k})\ket{e_j}\nonumber\\
     &=&\frac1{d^2}\sum_{k,j=1}^{d}
         \bra{e_j} T(\kettbra{f_k})\ket{e_j}\nonumber\\
     &=&\frac1{d^2}\sum_{j=1}^{d}
         \bra{e_j} T(\idty)\ket{e_j}\nonumber\\
     &=&\frac1d.
\end{eqnarray}
The same computation applies to $\tr\rho_1(T\otimes\id)(P_0)$, so
Alice's success probability is $1/d$, independently of her cheating
channel, and may hence be chosen to be arbitrarily small.


\subsection{A Decoherence Monster in Bob's Lab}
    \label{sec:monster}

In the proof of Th.~\ref{theo:nogo} we have shown that Alice has a
cheating strategy for any concealing protocol. Hence it is not
surprising that by weakening Alice's position, namely when
decoherence eliminates her favorite cheating option, bit
commitment protocols like those described in the previous section
become possible. But it may seem rather paradoxical that
decoherence acting on Bob's side, presumably further hampering the
weaker partner, can also lead to successful protocols.

Suppose that every morning, the cleaning service comes to Bob's
lab, unplugs all vacuum pumps, and restores what they take for
tidiness. Only classical records survive this procedure. When
Alice is convinced that she can rely on this, she might reassess
her demands on concealment, and the two might agree on a bit
commitment protocol, which under such circumstances is indeed both
concealing and binding. This example shows very clearly that the
entangled record introduced in the proof is essential.

The protocol we suggest relies on the distinction between the
local erasure of information and the destruction of quantum
correlations, as seen in a pair of channels demonstrating the
separation between ordinary operator norm and cb-norm in an
extreme way \cite{HLS+04}:
\begin{lemma}
    \label{lemma:randomize}
        Let $\varepsilon,\delta>0$. Then for sufficiently large $d$ there
        is a pair of channels $R,S:\bop{\Cx^d}\to\bop{\Cx^d}$ such that
        $\norm{R-S}\leq \varepsilon$ and $\cbnorm{R-S}\geq 2 - \delta$.
\end{lemma}
Since standard operator norm and cb-norm coincide for channels
with classical (Abelian) output space (cf.~Th.~3.9 in Paulsen's
text \cite{Pau02}), Lemma~\ref{lemma:randomize}
demonstrates a purely quantum-mechanical effect.\\
\\
{\bf Proof of Lemma~\ref{lemma:randomize}:} According to
Ref.~\cite{HLS+04}, a quantum channel $R \mathpunct : \bc{d}
\rightarrow \bc{d}$ is {\em $\varepsilon$-randomizing} iff
\begin{equation}
    \label{eq:monster01}
        \norm{R_{*}(\varrho) - S_{*}(\varrho)}_1 \leq
        \varepsilon\; \; \forall \; \; \varrho \in
        \bcstar{d},
\end{equation}
where $S$ denotes the completely depolarizing channel,
\begin{equation}
    \label{eq:monster02}
        S (e) = \frac{1}{d} \, \tr{\, e}
        \quad \Longleftrightarrow \quad S_{*} (\varrho) =
        \frac{\tr{\rho}}{d} \, \idty
\end{equation}
for all $e \in \bc{d}$ and $\varrho \in \bcstar{d}$, respectively.
Eq.~(\ref{eq:monster01}) implies the norm estimate $\norm{R - S}
\leq \varepsilon$, as required in Lemma~\ref{lemma:randomize}.

Hayden et al. show that for $d>\frac{10}{\varepsilon}$, such an
$\epsilon$-randomizing quantum channel can be obtained with high
probability from a random selection of at most $\mu := \lceil
\frac{134}{\varepsilon^2} d \log d \rceil$ unitary operators $\{
U_{i} \}_{i=1}^{\mu} \subset \bc{d}$,
\begin{equation}
    \label{eq:monster03}
        R(e) := \frac{1}{\mu} \, \sum_{i=1}^{\mu} \,
        U_{i}^{*} \, e \, U_{i}^{}.
\end{equation}
In striking contrast, exact randomization of quantum states (such
that $\varepsilon = 0$ in Eq.~(\ref{eq:monster01})) is known to
require an ancilla system of dimension $d^2 \gg \mu$
\cite{AMT+00}. However, this significant reduction in the size of
the ancilla space comes at a price: while the randomizing map $R$
erases local information, it preserves almost all the correlations
with a bystander system if $d$ is sufficiently large. In fact, it
is straightforward to show the upper bound
\begin{equation}
    \label{eq:monster04}
        \tracenorm{(R_{*} - S_{*}) \otimes \id \kettbra{\Omega}} \geq
        2 - \frac{2\mu}{d^2},
\end{equation}
where $\ket{\Omega} := \frac{1}{\sqrt{d}} \sum_{i=1}^{d} \ket{i,
i}$ again denotes the maximally entangled state on $\Cx^{d}
\otimes \Cx^{d}$. Eq.~(\ref{eq:monster04}) implies the bound
$\cbnorm{R-S} \geq 2 - \delta$, where $\delta := \frac{2\mu}{d^2}$
can be made as small as desired by choosing $d$ sufficiently
large. $\blacksquare$\\
\\
We can now set up a bit commitment protocol in which Bob initially
supplies a pure state $\ket{\psi}$ on a $d$-dimensional Hilbert
space $\hh_B$ according to the unitarily invariant Haar measure. There is only one round for Alice, requiring her to
send back a system with the same Hilbert space. Her honest
strategies are specified by a pair of channels
$T_k:\bop{\hh^k_A\otimes\hh_B}\to\bop{\hh_B}$ ($k=0,1$). We take
them to be locally coherent, i.e., implemented by a single
isometry $V_k:\hh_B\to\hh_A^k\otimes\hh_B$ each. Their
restrictions to Bob's side will be channels as provided by
Lemma~\ref{lemma:randomize}: $T_0^B(X)=V_0^*(\idty\otimes
X)V_0=R(X)$ and, similarly, $T_1^B=S$.

\begin{figure}
    \begin{center}
        \psfrag{p}[cc][cc]{$\ket{\psi}$}
        \psfrag{T}[cc][cc]{$T_{k}$}
        \psfrag{M}[cc][cc]{$M$}
        \psfrag{A}{{\large Alice}}
        \psfrag{B}{{\large Bob}}
        \psfrag{a}{$\hh_{A}^{k}$}
        \psfrag{b}{$\hh_{B}$}
        \psfrag{k}{$k \in \{0,1\}$}
        \psfrag{t}{time}
        \includegraphics[width=0.9\columnwidth]{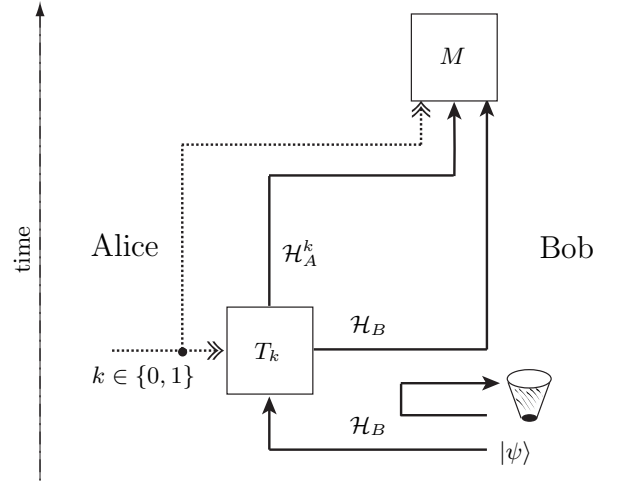}
        \caption{\label{fig:monster} {\it A quantum bit commitment
            protocol with local decoherence in Bob's lab. The rubbish bin symbolizes an
            entanglement-breaking channel acting on Bob's
            reference system. The figure shows the flow of quantum (solid) and classical (dashed)
            information if both
            Alice and Bob play honest. Alice controls all systems on the left-hand side of the
            figure, Bob those on the right-hand side. Time flows upwards. The protocol starts with
            Bob submitting some pure quantum state $\ket{\psi} \in \Cx^{d}$ to Alice, and ends with
            Bob's measurement $M$.}}
    \end{center}
\end{figure}

To reveal her commitment, Alice will later supply Bob with the
ancilla system $\hh_{A}^{k}$, alongside with the bit value $k$.
Bob will then verify Alice's claim with a projective measurement
on $V_{k} \ket{\psi}$, as illustrated in Fig.~\ref{fig:monster}.
Clearly, this protocol is perfectly {\em sound}, since Bob's
measurement will confirm the bit value $k$ with unit probability
if both parties have followed their honest strategies. The
protocol is $\frac{\varepsilon}{2}${\em -concealing}, provided the
Decoherence Monster strikes as planned, implementing some
entanglement-breaking channel \cite{HSR03} on Bob's reference
system. By definition, these are the channels $D \mathpunct :
\bhh{B} \rightarrow \bhh{B}$ such that $D \otimes \id (\varrho)$
is separable for any input state $\varrho$. Hence, these channels
are sometimes also called {\em separable}. In
Fig.~\ref{fig:monster}, the decoherence inflicted by $D$ is
indicated by the rubbish bin. We will show below that the maximal
probability difference Bob can detect by preparing suitable states
and making suitable measurements is then indeed just
$\norm{R-S}/2$.

To see that the protocol is binding note first, that Alice's usual
cheating strategy cannot work: If there were an operator $U$ such
that $(U\otimes\idty)V_0\approx V_1$ in norm, the two channels $R$
and $S$ could immediately be estimated to be cb-norm close, in
contradiction to the second property guaranteed by
Lemma~\ref{lemma:randomize}.

However, it is clearly not enough to argue that there is no {\em
universal} cheating strategy for Alice, which succeeds regardless
of Bob's input state. We need to rule out strategies which would
allow Alice to fool Bob's test in many cases, or with high
probability. In addition, we also have to show security for
arbitrary cheating strategies and, in particular, we have to make
certain that the reduction of Bob's lab capabilities by the
Decoherence Monster does not also give Alice a bit more freedom to
cheat. That is, in order to prove security we have to explain why
the coherent record makes a difference for Bob's ability to
distinguish the honest strategies, but not for his ability to
distinguish honest from cheating strategies in the opening phase.
This is the essence of the following
\begin{theo}
    \label{theo:monster}
        Let $\varepsilon >0$, $\delta >0$. Then for sufficiently
        large dimension $d$ the bit commitment protocol described
        above is perfectly sound, $\varepsilon$-concealing, and
        $\delta$-binding.
\end{theo}
In the proof of Th.~\ref{theo:monster} we will need to employ some
of the standard properties of distance measures for quantum states
and operations, which we collect here for reference. We start with
the well-known equivalence of the trace-norm distance and the
fidelity:
\begin{lemma}
    \label{lemma:normfidelity}
        Let $F(\varrho, \sigma) := {\rm tr} \,
        \sqrt{\sqrt{\varrho} \, \sigma \sqrt{\varrho}}$ denote the
        fidelity of two quantum states $\varrho$, $\sigma \in
        \bhstar$. We then have:
        \begin{equation}
            \label{eq:monster05}
                1 - F(\varrho, \sigma) \leq \frac{1}{2}
                \tracenorm{\varrho - \sigma} \leq
                \sqrt{1-F^2(\varrho, \sigma)}.
        \end{equation}
\end{lemma}
A proof of Lemma~\ref{lemma:normfidelity} can be found in Ch.~9.2
of \cite{NC00}.

The fidelity $F(\varrho, \sigma) = {\rm tr} \,\sqrt{\sqrt{\varrho}
\, \sigma \sqrt{\varrho}}$ is symmetric in its inputs and
unitarily invariant. It never decreases under quantum operations.
If $\varrho = \kettbra{\varphi}$ is pure, we have $F(\varphi,
\sigma) = \sqrt{\brakket{\varphi}{\sigma}{\varphi}}$. We will also
need the following lemma, which appears as Lemma~2 in \cite{SR01}:
\begin{lemma}
    \label{lemma:fidelity}
        For any two quantum states $\varrho$, $\sigma \in \bhstar$ we
        have:
        \begin{equation}
            \label{eq:monster06}
               \sup_{\omega \in \bhstar} \big \{ F^{2}(\varrho, \omega) +
               F^{2}(\sigma, \omega) \big \} = 1 + F(\varrho,
               \sigma).
        \end{equation}
\end{lemma}
We now proceed from quantum states to quantum operations: The {\em
channel fidelity} of a quantum channel $T \mathpunct : \bc{d}
\rightarrow \bc{d}$ is defined as
\begin{equation}
    \label{eq:monster07}
        F_{c}(T) := F^{2} \big ( \Omega, T \otimes \id (\kettbra{\Omega}) \big )
        = \brakket{\Omega}{T \otimes \id
        (\kettbra{\Omega})}{\Omega},
\end{equation}
where $\ket{\Omega} = \frac{1}{\sqrt{d}} \sum_{j=1}^{d} \ket{j}
\otimes \ket{j}$ is maximally entangled on $\Cx^{d} \otimes
\Cx^{d}$, as before. The channel fidelity $F_{c}(T)$ is a measure
for the quantum channel $T$ to preserve entanglement with a
bystander system, and is closely related to the {\em average
fidelity} of the channel $T$,
\begin{equation}
    \label{eq:monster08}
        \overline{F}(T) := \int
        \brakket{\psi}{T(\kettbra{\psi})}{\psi} \; d\psi,
\end{equation}
where the integral is over the normalized Haar measure:
\begin{lemma}
    \label{lemma:average}
        For any quantum channel $T \mathpunct :
        \bc{d} \rightarrow \bc{d}$, we have:
        \begin{equation}
            \label{eq:monster09}
                \overline{F}(T) \geq F_{c}(T) \geq \overline{F}(T)
                - \frac{1}{d}.
        \end{equation}
\end{lemma}
The proof of Lemma~\ref{lemma:average} is immediate from the
relation \cite{HHH99,Nie02}
\begin{equation}
    \label{eq:monster10}
        \overline{F}(T) = \frac{d F_{c}(T) + 1}{d+1}. \;
        \blacksquare
\end{equation}
In the protocol described above we grant the Decoherence Monster
the freedom to apply an arbitrary {\em entanglement-breaking}
quantum channel on Bob's bystander system. Any such channel $D
\mathpunct : \bhh{B} \rightarrow \bhh{B}$ can be decomposed
\cite{HSR03} as $D_{} = D_{1} \circ D_{2}$, where
\begin{align}
    \label{eq:monster11}
        D_{1} & \mathpunct : \; \cC_{X} \rightarrow \bhh{B} \qquad
        {\rm and}\\
    \label{eq:monster11a}
        D_{2} & \mathpunct : \; \bhh{B} \rightarrow \cC_{X},
\end{align}
and $\cC_{X}$ denotes the Abelian algebra of the complex-valued
functions on the finite set $X$ (with $\abs{X}$ elements). In
other words, any entanglement-breaking channel can be thought of
as being built from a measurement channel $D_{1}$, with resulting
classical output system $\cC_{X}$, followed by a re-preparation
$D_{2}$. (Note that Eqs.~(\ref{eq:monster11}) and
(\ref{eq:monster11a}) describe the channels $D_{k}$ in the
Heisenberg picture, so the direction of arrows is inverted, cf.
Sec.~\ref{sec:observables}.)

In order to confirm $\varepsilon$-concealment of the monster
protocol, we will need to show that any such entanglement-breaking
channel $D$ renders Bob's bystander system useless for the
analysis of Alice's actions:

\begin{lemma}
    \label{lemma:separable}
        For any linear map $L \mathcal : \bh \rightarrow \bk$ and
        any entanglement-breaking channel $D \mathcal : \bhh{1}
        \rightarrow \bkk{1}$,
        \begin{equation}
            \label{eq:monster11b}
                \norm{L \otimes D} = \norm{L}.
        \end{equation}
\end{lemma}
Since entanglement-breaking channels have a decomposition $D_{1}
\circ D_{2}$ with an intermediate classical system $\cC_{X}$, it
will turn out sufficient to verify this property for the noiseless
classical channel $\id_{X}$:

\begin{lemma}
    \label{lemma:classical}
        For any linear map $L \mathcal : \bh \rightarrow \bk$ and
        any classical observable algebra $\cC_{X}$,
        \begin{equation}
            \label{eq:monster11c}
                \norm{L_{} \otimes \id_{X}} = \norm{L}.
        \end{equation}
\end{lemma}
{\bf Proof of Lemma~\ref{lemma:classical}:} For any $a \in \bh$ we
have,
\begin{equation}
    \label{eq:monster11d}
        \norm{L(a)} = \norm{L_{} \otimes \id_{X} \, (a_{} \otimes
        \idty_{X})} \leq \norm{L_{} \otimes \id_{X}} \,
        \norm{a_{}},
\end{equation}
which shows that $\norm{L_{}} \leq \norm{L_{} \otimes \id_{X}}$.

For the converse implication, note that any classical-quantum
state $\varrho$ on $\bk \otimes \cC_{X}$ is of the form
\begin{equation}
    \label{eq:monster11e}
        \varrho = \sum_{x=1}^{\abs{X}} p_{x} \, \varrho_{x}
        \otimes \ketbra{x}{x},
\end{equation}
where $\{p_{x}\}_{x=1}^{\abs{X}}$ is a classical probability
distribution, $\{\varrho_{x}\}_{x=1}^{\abs{X}}$ is a set of
quantum states on $\bk$, and $\{ \ket{x} \}_{x=1}^{\abs{X}}$
denotes an orthonormal basis for $\C^{\abs{X}}$ (cf.~Prop.~2.2.4
in \cite{Key02}). We may now estimate,
\begin{equation}
    \label{eq:monster11f}
        \begin{split}
            \tracenorm{(L_{*} \otimes \id_{X}) \varrho} & \leq
            \sum_{x=1}^{\abs{X}} p_{x} \, \tracenorm{L_{*}
            (\varrho_{x}) \otimes \ketbra{x}{x}}\\
            & = \sum_{x=1}^{\abs{X}} p_{x} \, \tracenorm{L_{*}
            (\varrho_{x})}\\
            & \leq \sum_{x=1}^{\abs{X}} p_{x} \, \norm{L} = \norm{L},
        \end{split}
\end{equation}
and hence $\norm{L_{} \otimes \id_{X}} \leq \norm{L_{}}$, as
claimed. $\blacksquare$\\
\\
{\bf Proof of Lemma~\ref{lemma:separable}:} Choosing $a \in \bh$,
we immediately have
\begin{equation}
    \label{eq:monster11g}
        \begin{split}
            \norm{L(a)} & = \norm{L(a) \otimes \idty_{\kk_{1}}}\\
            & = \norm{(L \otimes D) \, (a \otimes
            \idty_{\hh_{1}})}\\
            & \leq \norm{L \otimes D} \, \norm{a \otimes
            \idty_{\hh_{1}}} \\
            & = \norm{L \otimes D} \, \norm{a},
        \end{split}
\end{equation}
implying $\norm{L} \leq \norm{L \otimes D}$.

For the converse implication, let $D_{} = D_{1} \circ D_{2}$ be a
decomposition as in Eqs.~(\ref{eq:monster11}) and
(\ref{eq:monster11a}) above. We may then estimate,
\begin{equation}
    \label{eq:monster11h}
        \begin{split}
            \norm{L \otimes D} & = \norm{L \otimes (D_{1} \circ
            D_{2})}\\
            & = \norm{(\id_{\kk} \otimes D_{1}) \, (L \otimes
            \id_{X}) \, (\id_{\hh} \otimes D_{2})}\\
            & \leq \norm{\id_{\kk} \otimes D_{1}} \,
            \norm{L \otimes \id_{X}} \, \norm{\id_{\hh} \otimes
            D_{2}}\\
            & \leq \cbnorm{D_{1}} \, \norm{L \otimes \id_{X}} \,
            \cbnorm{D_{2}} = \norm{L},
        \end{split}
\end{equation}
where in the last step we have used Lemma~\ref{lemma:classical}
and the fact that $\cbnorm{T}=1$ for any channel $T$
(cf.~Sec.~\ref{sec:distance}). $\blacksquare$\\
\\
We now have all the tools at hand to complete the\\
\\
{\bf Proof of Th.~\ref{theo:monster}:} Soundness of the protocol
is clear. Setting $L:= R-S$ in Lemma~\ref{lemma:separable},
$\varepsilon$-concealment follows immediately from
Lemma~\ref{lemma:randomize}.

Thus, it only remains to show that the protocol is
$\delta$-binding. As a warm-up exercise, let us first exclude the
possibility of Alice committing to the bit value $k$ in the
commitment phase, and then announcing the bit $1-k$ in the opening
phase. This is sometimes called {\em passive cheating}.

If Bob has initially supplied the pure state $\ket{\psi} \in
\Cx^{d}$, the probability of successfully passing Bob's projective
measurement in such a scenario is $P(\psi) := \abs{\braket{V_{0}
\psi}{V_{1} \psi}}^{2}$, resulting in the overall cheating
probability
\begin{eqnarray}
    \label{eq:monster12}
        P &:=& \int P(\psi) \; d\psi \nonumber\\
        &=& \int \brakket{\psi}{V_{0}^{*}
        V_{1}^{} (\kettbra{\psi}) V_{1}^{*} V_{0}^{}}{\psi} \;
        d\psi \nonumber \\
        &\overset{(\ref{eq:monster08})}{=}& \overline{F}(V_{0}^{*} V_{1}^{}).
\end{eqnarray}
For $\delta$ and $d$ as in Lemma~\ref{lemma:randomize}, we then
have the estimate
\begin{eqnarray}
    \label{eq:monster13}
            2 - \delta & \overset{(\ref{eq:monster04})}{\leq} &
            \tracenorm{T_{0*}^{B} \otimes \id (\kettbra{\Omega}) -
            T_{1*}^{B} \otimes \id (\kettbra{\Omega})} \nonumber\\
            & \leq & \tracenorm{(V_{0}^{} \otimes \idty)
            \kettbra{\Omega} (V_{0}^{*} \otimes \idty) \nonumber \\
            && \qquad \qquad \qquad - (V_{1}^{} \otimes
            \idty) \kettbra{\Omega} ( V_{1}^{*} \otimes \idty)} \nonumber \\
            & \overset{(\ref{eq:monster05})}{\leq} & 2 \, \sqrt{1 -
            F^{2} (V_{0} \otimes \idty \ket{\Omega}, V_{1} \otimes \idty
            \ket{\Omega})} \nonumber \\
            & \overset{(\ref{eq:monster07})}{=} & 2 \,
            \sqrt{1 - F_{c}(V_{0}^{*} V_{1}^{})} \nonumber \\
            & \overset{(\ref{eq:monster09})}{\leq} & 2 \,
            \sqrt{1 - \overline{F}(V_{0}^{*} V_{1}^{}) + \frac{1}{d}} \nonumber \\
            & \overset{(\ref{eq:monster12})}{=} & 2 \,
            \sqrt{1 - P + \frac{1}{d}},
\end{eqnarray}
where in the second step we have used that the trace-norm cannot
increase under the partial trace operation \cite{NC00}. From
Eq.~(\ref{eq:monster13}) we conclude that
\begin{equation}
    \label{eq:monster14}
        P \leq \frac{1}{d} + \delta.
\end{equation}
Since the right side of Eq.~(\ref{eq:monster14}) can be made as
small as desired by stepping up the dimension, this gives the
desired upper bound on Alice's probability of successfully passing
Bob's test.

\begin{figure}
    \begin{center}
        \psfrag{p}[cc][cc]{$\ket{\psi}$}
        \psfrag{T}[cc][cc]{$T^{\sharp}$}
        \psfrag{S}[cc][cc]{$T^{\sharp}_{k}$}
        \psfrag{M}[cc][cc]{$M$}
        \psfrag{A}{{\large Alice}}
        \psfrag{B}{{\large Bob}}
        \psfrag{a}{$\hh_{A}^{k}$}
        \psfrag{c}{$\hh_{\sharp}$}
        \psfrag{b}{$\hh_{B}$}
        \psfrag{k}{$k \in \{0,1\}$}
        \psfrag{t}{time}
       \includegraphics[width=0.9\columnwidth]{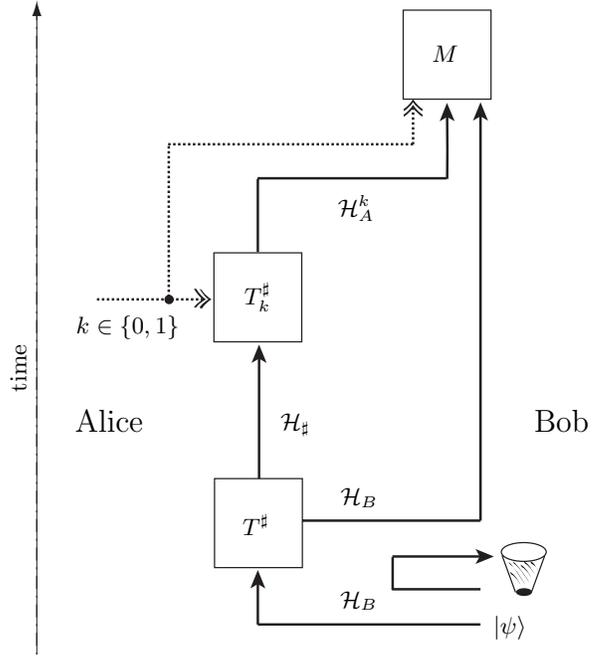}
        \caption{\label{fig:cheat} {\it Alice's cheating strategy
            consists of applying some quantum channel $T^{\sharp}$ in the commitment phase, and then
            another quantum channel $T^{\sharp}_{k}$ to commit to the bit value $k \in \{0,1\}$ only
            before the opening. Her goal is to pass Bob's projective measurement $M$.}}
    \end{center}
\end{figure}

So far we have only proven bindingness against passive cheating
attacks. As illustrated in Fig.~\ref{fig:cheat}, Alice's most
general attack consists of applying some quantum channel
$T^{\sharp} \mathpunct : \bop{\hh_{\sharp}} \otimes \bop{\hh_{B}}
\rightarrow \bop{\hh_{B}}$ during the commitment phase,
independently of the bit value $k \in \{0,1\}$. She will send a
$d$-dimensional quantum system $\hh_{B}$ to Bob without having
committed to either bit. Only before the opening will she then
decide on a bit value $k$, apply a corresponding quantum channel
$T^{\sharp}_{k} \mathpunct : \bop{\hh_{A}^{k}} \rightarrow
\bop{\hh_{\sharp}}$ on her remaining system, and hope to pass
Bob's projective measurement.

Assuming that Alice is a not prejudiced towards either bit, the
probability of passing Bob's test is then $P := \frac{1}{2} P_{0}
+ \frac{1}{2}P_{1}$, where for $k \in \{0,1\}$ we set
\begin{equation}
    \label{eq:monster15}
        P_{k} := \int \brakket{V_{k} \psi}{(T_{k*}^{\sharp}
        \otimes \id_{B}) T^{\sharp}_{*} (\kettbra{\psi})}{V_{k}
        \psi} \; d\psi.
\end{equation}
This probability can be bounded as follows:
\begin{eqnarray}
    \label{eq:monster16}
        P_{k} & = & \int \brakket{\psi}{V_{k}^{*}(T_{k*}^{\sharp}
        \otimes \id_{B}) T^{\sharp}_{*} (\kettbra{\psi}) V_{k}^{}}
        {\psi} \; d\psi \nonumber \\
        & \overset{(\ref{eq:monster08})}{=} & \overline{F} \big ( V_{k}^{*}
        (T_{k*}^{\sharp} \otimes \id_{B}) T^{\sharp}_{*}
        V_{k}^{} \big ) \nonumber \\
        & \overset{(\ref{eq:monster09})}{\leq} & F_{c} \big ( V_{k}^{*}
        (T_{k*}^{\sharp} \otimes \id_{B}) T^{\sharp}_{*}
        V_{k}^{} \big ) + \frac{1}{d}\\
        & \overset{(\ref{eq:monster07})}{=} & F^{2} \big ( V_{k} \otimes
        \idty_{B'} \ket{\Omega}, \, (T_{k*}^{\sharp} \otimes
        \id_{B} \otimes \id_{B'}) \nonumber \\
        && \qquad \qquad \qquad \qquad (T^{\sharp}_{*} \otimes
        \id_{B'})(\kettbra{\Omega}) \big ) + \frac{1}{d} \nonumber \\
        & \leq & F^{2} \big ( T_{k*}^{B} \otimes \id_{B'}
        (\kettbra{\Omega}), \nonumber \\
        && \qquad \qquad \qquad \qquad {\rm tr}_{\hh_{\sharp}} T_{*}^{\sharp}
        \otimes \id_{B'} (\kettbra{\Omega}) \big ) + \frac{1}{d} \nonumber,
\end{eqnarray}
where in the final step we have used the monotonicity of the
fidelity under the partial trace operation. Combining this
estimate with Lemma~\ref{lemma:fidelity} and
Eq.~(\ref{eq:monster04}) then immediately yields the bound
\begin{eqnarray}
    \label{eq:monster17}
        P &\leq& \frac{1}{2} + \frac{1}{d} + \frac{1}{2} \,
        F \big ( R \otimes \id_{B'}(\kettbra{\Omega}), S \otimes \id_{B'}
        (\kettbra{\Omega}) \big ) \nonumber \\
        & \leq&  \frac{1}{2} + \frac{1}{d} + \frac{1}{2}
        \sqrt{\delta}.
\end{eqnarray}
The right-hand side can be brought as close to $\frac{1}{2}$ as
desired by stepping up the dimension. Resubstituting $\frac{1}{d}
+ \frac{1}{2} \sqrt{\delta} \mapsto \delta$, the protocol is
$\delta$-binding. This concludes the proof of
Th.~\ref{theo:monster}. $\blacksquare$


\subsection{Decoherence in the Transmission Line}
    \label{sec:transmission}

While noise in the transmission line is generally considered a
nuisance, and coding theorists have designed elaborate error
correcting codes to cope with it, Wyner \cite{Wyn75} was the first
to realize that noise may sometimes be beneficial for
cryptographic applications
--- in his case for key distribution. Cr\'epeau and Kilian
\cite{CK88} have later shown that classical noisy channels may
also be employed to establish secure bit commitment. Their results
have subsequently been improved in \cite{Cre97,DKS99}. Recently
Winter {\it et al.} \cite{WNI03} have considered the asymptotic
version of string commitment and have obtained a single-letter
expression for the commitment capacity of a classical noisy
channel. Their results show that any nontrivial noisy channel can
be used to establish secure bit commitment. The theorem can be
extended to so-called {\em classical-quantum} channels. But it
remains an open question whether fully quantum channels can also
be useful for bit commitment.

Misaligned spatial reference frames can also effectively act as a
noisy channel, and facilitate secure bit commitment. An example
for a secure protocol was recently given by Harrow {\it et al.}
\cite{HOT05}.


\section{Summary and Discussion}
    \label{sec:discussion}

In summary, we have presented a general framework for two-party
cryptographic protocols and have shown that secure quantum bit commitment
is impossible within that framework --- by giving explicit bounds
on the degree of concealment and bindingness that can be
simultaneously achieved in any given protocol. Our proof covers
protocols on finite or infinite dimensional Hilbert spaces with
any number of rounds in each of the commitment, holding, and opening
phases. In contrast to earlier proofs, we do not assume the
receiver to be bound to a publicly known strategy. Thus, our
strengthened no-go result also covers the anonymous state
protocols that have been repeatedly suggested as a way to
circumvent the standard no go arguments. If the receiver's
strategy is fixed and common knowledge, our bounds coincide with
those obtained by Spekkens and Rudolph \cite{SR01}, and hence the
standard no-go proof is recovered in that case.

Our formulation of the no-go proof contains an explicit treatment
of the classical information flow, possibly of independent
interest for other cryptographic applications. As a consequence,
the framework directly applies to the purely classical setting, in
which no quantum information is exchanged and all local Hilbert
spaces are one-dimensional. Note however, that in order to cheat
with a sneak flip operation as described in the proof of
Th.~\ref{theo:nogo}, in general Alice will nevertheless need to
apply a quantum operation. This is so because the commutant of
Bob's classical system is usually a hybrid containing both
classical and quantum parts. Hence, a classical protocol embedded
in a quantum world allows Alice to cheat, but the fully classical
no-go proof is not directly recovered.

We emphasize that in our setup, Alice and Bob may draw on an
unlimited supply of certified classical or quantum correlations,
in the form of an arbitrary shared initial state $\rho_0$, and yet
secure quantum bit commitment remains impossible. This is in striking
contrast to quantum coin tossing: starting with a maximally
entangled qubit state and measuring in a fixed basis, Alice and Bob can
obviously implement a perfectly fair and secure coin tossing
protocol\footnote{We would like to thank the referee for
clarifying this point.}.

In the second part of the paper, we have analyzed quantum bit
commitment protocols relying on decoherence. We have presented a
new such protocol in which provably secure bit commitment is
guaranteed through an entanglement-breaking channel in the receiver's
lab. The protocol relies on the separation between local erasure
of information and the destruction of correlations, which is a
purely quantum mechanical effect.

In accordance with most of the literature, throughout this work we
have restricted the discussion to quantum bit commitment protocols
in which concealment is guaranteed for all branches of the
communication tree. This is sometimes called {\em strong} bit
commitment, in order to distinguish it from a weaker form in which
Bob may possibly learn the value of the bit --- as long as Alice
receives a message stating that the bit value has been disclosed.
{\em Weak} bit commitment protocols have been analyzed by Hardy
and Kent \cite{HK04}, and independently by Aharonov {\em et al.}
\cite{ATV+00}. Such protocols are sufficient whenever bit
commitment is only part of a larger cryptographic environment, and the
value of the bit itself does not reveal any useful information. In
particular, secure weak bit commitment protocols could be applied
to implement quantum coin tossing. However, weak bit commitment is
likewise impossible, with identical bounds on the concealment and
bindingness. The no-go proof follows our analysis for strong bit
commitment in this paper, but the concealment condition now only
has to be guaranteed for a subtree, and hence for a subchannel.
Alice then finds a sneak flip operation from a version of
Stinespring's continuity theorem for subnormalized quantum
channels \cite{KSW07}.


\section{Appendix: Language and Notations}
    \label{sec:notation}

This appendix contains the necessary background on observables,
states and quantum channels, as well as on direct sums and their
role for the description of algebraically encoded classical
information. We restrict the discussion to the basics, and refer
to the textbook of Bratteli and Robinson \cite{BR87} and Keyl's survey article
\cite{Key02} for a more complete presentation.


\subsection{Observables, States, and Quantum Channels}
    \label{sec:observables}

The statistical properties of quantum systems are characterized by
spaces of operators on a Hilbert space $\hh$: The observables of
the system are given by bounded linear operators on $\hh$, written
$\bh$. This is the prototype of a $C^{*}$-algebra and is usually
called the {\em observable algebra} of the system. The physical
states are then those positive linear functionals $\omega \mathpunct
: \bh \rightarrow \C$ that satisfy the normalization condition
$\omega (\idty) = 1$. We restrict our discussion to
finite-dimensional Hilbert spaces, for which all linear operators
are bounded and every linear functional $\omega$ can be expressed
in terms of a trace-class operator $\varrho_{\omega} \in \bhstar$
such that $\omega(A) = \tr{\varrho_{\omega} A}$ for all $A \in
\bh$. The normalization of the functional $\omega$ than translates
into the condition $\tr{\varrho_{\omega}} = 1$. The physical
states can thus be identified with the set of normalized density
operators $\varrho
\in \bhstar$.\\

A {\em quantum channel} $T$ which transforms input systems
described by a Hilbert space $\hh_{A}$ into output systems
described by a (possibly different) Hilbert space $\hh_{B}$ is
represented (in the Heisenberg picture) by a completely positive
and unital map $T \mathpunct: \bhh{B} \rightarrow \bhh{A}$. By
unitality we mean that $T(\idty_{B}) = \idty_{A}$, with the
identity operator $\idty_{X} \in \bhh{X}$. Complete positivity
means that $\id_{\nu} \otimes T$ is positive for all $\nu \in \N$,
where $\id_{\nu}$ denotes the identity operation on the $(\nu
\times \nu)$ matrices.

The physical interpretation of the quantum channel $T$ is the
following: when the system is initially in the state $\varrho \in
\bhhstar{A}$, the expectation value of the measurement of the
observable $B \in \bhh{B}$ at the output side of the channel is
given in terms of $T$ by $\tr{\, \varrho \: T(B)}$. Unitality
provides the normalization, while complete positivity guarantees
that all expectation values remain positive even if the channel is
only part of a larger network.

Alternatively, we can focus on the dynamics of the states and
introduce the dual map $T_{*} \mathpunct : \bhhstar{A} \rightarrow
\bhhstar{B}$ by means of the duality relation
\begin{equation}
    \label{eq:dual}
        \tr{\, T_{*}(\varrho) \, B} = \tr{\varrho \, T(B)} \quad
        \forall \; \; \varrho \in \bhhstar{A}, B \in \bhh{B}.
\end{equation}
$T_{*}$ is a completely positive and trace-preserving map and
represents the channel in the {\em Schr\"odinger picture}, while $T$
provides the {\em Heisenberg picture} representation. For
finite-dimensional systems, the Schr\"odinger and the Heisenberg
picture provide a completely equivalent description of physical
processes. The interconversion is always immediate from
Eq.~(\ref{eq:dual}).

\subsection{Direct Sums and Quantum-Classical Hybrid Systems}
    \label{sec:sums}

Our general description of bit commitment protocols includes a
full treatment of the classical and quantum information flow. As explained in
Section~\ref{sec:formal}, the nodes of the communication tree
correspond to the classical information accumulated in the course
of the protocol. Direct sums are a convenient way to encode this
information in the observable algebras: For a finite collection of
observable algebras $\{\cA_x \}_{x \in X}$, the direct sum algebra
\begin{equation}
    \label{eq:sums01}
        \bigoplus_{x=1}^{X} \cA_x := \{ \bigoplus_{x=1}^{X} A_{x} \; \mid
        \; A_{x} \in \cA_x \}
\end{equation}
represents the physical situation in which the system under
consideration is described by an observable algebra $\cA_x$ if the
classical information $x \in X$ has been accumulated. Sums and
products as well as adjoints in this algebra are defined
component-wise, i.\,e.,
\begin{eqnarray}
    \label{eq:sums02}
        \bigoplus_x A_x + \bigoplus_x B_x &:=& \bigoplus_x \big ( A_x + B_x \big
        )\\
        \bigoplus_x A_x \cdot \bigoplus_x B_x &:=& \bigoplus_x \big (A_x \cdot
        B_x \big )\\
        \alpha \cdot \bigoplus_x A_x &:=& \bigoplus_x \big ( \alpha
        \cdot A_x \big ) \\
        \big ( \bigoplus_x A_{x}^{} \big )^* &:=& \bigoplus_x
        A_{x}^{*}
\end{eqnarray}
for all operators $A_x, B_x \in \cA_x$, and coefficients $\alpha
\in \C$. It is straightforward to verify that with these
definitions $\oplus_x \cA_x$ is indeed an algebra with identity
$\idty = \oplus_x \idty_x$, where for each $x \in X$ $\idty_x$
denotes the identity in $\cA_x$. The norm on $\oplus_x \cA_x$ is
given by
\begin{equation}
    \label{eq.sums03}
        \norm{\oplus_x A_x} := \max_{x \in X} \norm{A_x}.
\end{equation}

If $\cA_x = \bhh{x}$ for a collection of Hilbert spaces $\{\hh_x
\}_{x=1}^{X}$, then $\oplus_{x} \bhh{x} \subset \mathcal{B} \big (
\oplus_x \hh_x \big )$. The physical states on such a system are
of the form $\oplus_x p_x \varrho_x$, where $\varrho_x \in
\bhhstar{x}$ are states on the component algebras and $\{p_x
\}_{x=1}^{X}$ is a classical probability distribution.

As explained in Section~\ref{sec:formal}, in our formulation of
the bit commitment protocol the component algebras $\cA_x$ will
usually be tensor products of observable algebras in Alice's and
Bob's lab, respectively: $\cA_x = \cA_x(a) \otimes \cB_x(b)$. The
local algebras $\cA_x(a)$ and $\cB_x(b)$ could be full matrix
algebras, or could themselves be direct sums, representing local
classical information available to Alice or Bob exclusively. The
strategic operations that are performed by Alice and Bob are
described by channels acting on these direct sum algebras. In the
Heisenberg picture, these channels are completely positive unital
maps $T:\cA\to\bop\hh$ with $\cA=\oplus_x\cA_x$. Their
interpretation is easily seen from Stinespring's representation
(Prop.~\ref{stinespring}): There exists a Hilbert space $\kk$, an
isometry $V \mathpunct : \hh\to\kk$ as well as a representation
$\pi$ of $\cA$ such that  $T(A)=V^*\pi(A)V$ holds.  For each $x\in
X$, the identity operator of the direct summand $\cA_x$ is a
projection $P_x$ in $\cA$ that commutes with all operators in
$\cA$. These projections generate an abelian subalgebra $C(\cA)$
called the {\em center} of $\cA$. Since $\pi$ is a
*-representation and therefore respects the product of operators,
$\pi(P_x)$ projects onto the subspace $\pi(P_x)\kk=:\kk_x$, which
is invariant under the action of all represented operators
$\pi(\cA)$. Hence we obtain for every $x$ a representation of
$\cA$ on $\kk_x$ according to
\begin{equation}
    \label{eq:appendix10}
        \pi_x(A):=\pi(P_x)\pi(A)\pi(P_x)=\pi(A)\pi(P_x).
\end{equation}
Since each direct summand $\cA_x=\bop{\hh_x}$ is a full matrix
algebra, the Hilbert spaces $\kk_x$ can be chosen to be of the
form $\kk_x=\hh_x\otimes\mm_x$ with appropriate multiplicity
spaces $\mm_x$. The representation $\pi_x$ is then given by
$\pi_x(\oplus_xA_x)=A_x\otimes\idty_{\mm_x}$. In terms of the
representations $\pi_x$, the action of the channel $T$ on an
operator $A$ can be written as
\begin{equation}
    T(A)=\sum_{x \in X} V^*\pi_x(A)V \; .
\end{equation}

How is this kind of representation interpreted in operational
terms? We first have a look at measurement operations in the
Heisenberg picture. Usually a measurement operation is described
by a {\em positive operator valued measure} (POVM), i.e., a
collection
\begin{equation}
    \{M_x\in\bop\kk \, |\, 0\leq M_x\leq \idty,
    \mbox{$\sum_x M_x=\idty$} \} \; .
\end{equation}
The set $X$ is interpreted as the set of possible measurement
outcomes. In the Heisenberg picture, this corresponds to a
completely positive normalized map $M$ from the abelian algebra
$\oplus_x\Cx=\Cx^X$ into $\bop\kk$. Namely, the operator
$f\in\Cx^X$ is mapped to $M(f)=\sum_x M_x f_x$. Hence, measurement
operations are a special class of channels on direct sum algebras,
where each summand is chosen to be one-dimensional, $\cA_x=\Cx$.
Thus, if we restrict the channel $T$ to the center $C(\cA)$, which
is isomorphic to $\Cx^X$, then we obtain a measurement operation
whose corresponding POVM is given by the operators $\{
V^*\pi(P_x)V|x\in X \}$. To verify this, we evaluate $T$ on a
central element $C\in C(\cA)$,
\begin{equation}
    T(C)=T\left(\sum_x C_x P_x\right)=\sum_x V^*\pi(P_x)V C_x\; ,
\end{equation}
where central elements $C$ are expressed as linear combinations of
the projections $P_x$, i.e., $C=\sum_x C_xP_x$ with $C_x\in\Cx$.
This justifies the following interpretation: The quantum system
under investigation is described by the observable algebra $\cA_x$
if the measurement results in the outcome $x\in X$. In other
words, the direct sum operation can be seen as a ``logical XOR"
composition of quantum systems --- in contrast to the tensor
product, which corresponds to the ``logical AND".

Coming back to the bit commitment protocol, the nodes of the
communication tree then in fact have a natural interpretation as
outcomes of a measurement process returning a history of
communicated decisions, which are given by the unique path in the
tree starting at its root and ending at $x \in X$.

\subsection{Distance Measures}
    \label{sec:distance}

In order to evaluate the concealment and bindingness conditions in
a quantum bit commitment protocol we need to measure the distance
between two quantum channels or two quantum states: Assume two
channels $T_{1}$ and $T_{2}$ with common input and output algebras
$\cA$ and $\cB$, respectively. Since these $T_{i}$ are (in
Heisenberg picture) operators between normed spaces $\cB$ and
$\cA$, the natural choice to quantify their distance is the {\em
operator norm},
\begin{equation}
    \label{eq:distance01}
        \norm{T_{1} - T_{2}} := \sup_{B \neq 0} \frac{\norm{T_{1}(B) -
        T_{2}(B)}}{\norm{B}}.
\end{equation}
The norm distance Eq.~(\ref{eq:distance01}) has a neat operational
characterization: it is twice the largest difference between the
overall probabilities in two statistical quantum experiments
differing only in replacing one use of $T_{1}$ with one use of
$T_{2}$.

However, we also want to allow for more general experiments, in
which the two channels are only applied to a sub-system of a
larger system. This requires {\em stabilized} distance measures
\cite{GLN04}, and naturally leads to the so-called {\em norm of
complete boundedness} (or {\em cb-norm}, for short) \cite{Pau02}:
\begin{equation}
    \label{eq:distance02}
        \cbnorm{T_{1} - T_{2}} := \sup_{\nu \in \N}
        \norm{ \id_{\nu} \otimes ( T_{1}
        - T_{2} )},
\end{equation}
where $\id_{\nu}$ again denotes the {\em ideal} (or {\em
noiseless}) channel on the $(\nu \times \nu)$-matrices. Useful
properties of the cb-norm include {\em multiplicativity}, i.\,e.,
$\cbnorm{T_{1} \otimes T_{2}} = \cbnorm{T_1} \, \cbnorm{T_2}$, and
{\em unitality:} $\cbnorm{T} = 1$ for any channel $T$.

Obviously, $\cbnorm{T} \geq \norm{T}$ for every linear map $T$. If
either the input or output space is a classical system, we even
have equality: $\cbnorm{T} = \norm{T}$ (cf. Ch.~3 in
\cite{Pau02}). Fully quantum systems generically show a
separation between these two norms.\\

States are channels with one-dimensional input space, $\C$. Since
this is a classical system, there is no need to distinguish
between stabilized and non-stabilized distance measures. The
so-called {\em trace norm} $\tracenorm{\varrho} = {\rm tr \,}
\sqrt{\varrho^{*} \varrho^{}}$ is frequently employed to evaluate
the distance between two density operators. The trace norm
difference $\tracenorm{\varrho - \sigma}$ is equivalent to the
{\em fidelity} $F(\varrho,\sigma) := {\rm tr} \,
\sqrt{\sqrt{\varrho} \, \sigma \sqrt{\varrho}}$ (cf.
Lemma~\ref{lemma:normfidelity}).

For any linear operator $T$ the operator norm $\norm{T}$ equals
the norm of the Schr\"odinger adjoint $T_{*}$ on the space of
trace class operators, i.\,e.,
\begin{equation}
    \label{eq:distance05}
        \norm{T_{}} = \sup_{\tracenorm{\varrho} \leq 1} \;
        \tracenorm{T_{*} (\varrho)}
\end{equation}
(cf. Ch.~VI of \cite{RS80} and Section~2.4 of \cite{BR87} for
details), which is the usual way to convert norm estimates from
the Heisenberg picture into the Schr\"odinger picture and vice
versa. For states $T_{*} = \varrho$, the operator norm then indeed
just coincides with the trace norm: $\norm{T} = \tracenorm{T_{*}}
= \tracenorm{\varrho}$.



\begin{acknowledgments}
    We would like to thank A. Winter, A. S. Holevo, M. Mosca,
    A. Kent, J. Oppenheim, M. Christandl, R. Renner, R. K\"onig,
    R. Colbeck, L. Ioannou, M. Keyl, and C. D\"oscher for
    fruitful and committed discussions that generated both heat
    and light. A. Harrow generously shared his insight on bit
    commitment with local decoherence in Bob's lab. GMD
    acknowledges extensive and detailed discussions with H. Yuen,
    whose work essentially stimulated the present one. GMD
    also acknowledges interesting discussions with H. K. Lo
    and with R. Spekkens, and with C. Bennett for clarifying
    the general philosophy and attitude at the basis of the
    previous impossibility proofs. We are indebted to an anonymous
    referee for a very careful reading of the manuscript
    and several valuable suggestions, in particular on protocols
    with trusted shared classical or quantum correlations
    and protocols with infinitely many rounds.\\
    \\
    DK is grateful for financial support from the European Union
    project RESQ and the German Academic Exchange Service (DAAD).
    GMD acknowledges financial support from Ministero Italiano
    dell'Universit\`a e della Ricerca (MIUR) through PRIN
    (bandi 2001, 2003, and 2005) and FIRB (bando 2001), and from the
    US Defence Advanced Research Project Agency under grant
    F30602-01-2-0528. Part of the work was done at the Max Planck
    Institute of Complex Systems in Dresden during the International
    Summer School on Quantum Information in Sept.~2005.
\end{acknowledgments}


\end{document}